\documentclass{aa}
\usepackage[english]{babel}
\usepackage[varg]{txfonts}

\usepackage[usenames,dvipsnames]{xcolor}

\graphicspath{{./}{figures/}}

\usepackage{epsfig}
\usepackage{graphicx,natbib}
\usepackage{amsmath,amssymb}
\usepackage{hyperref}
\usepackage{wasysym}
\usepackage{natbib}
\usepackage{ulem}
\usepackage{cancel}
\usepackage{subfigure}
\usepackage{gensymb}

\hypersetup{
    colorlinks=true,
    citecolor=blue,
    linkcolor=red,
    urlcolor=black
    }

\begin{document}

\def\nat{Nature }
\def\apj{Astrophys. J. }
\def\apjs{Astrophys. J., Suppl. Ser. }
\def\apjl{Astrophys. J., Lett. }
\def\apss{Astrophys. and Space Science}
\def\aap{A\&A}
\def\mnras{MNRAS}
\def\jgr{J. Geo. Reas.}
\def\zat{Zeitschrift f\"ur Astrophysik}

\def\nh{n_\mathrm{H}}
\def\mh{m_\mathrm{H}}
\def\nho{n_{\mathrm{H}0}}
\def\cc{\mathrm{cm}^{-3}}
\def\gcc{\mathrm{g}~\mathrm{cm}^{-3}}
\def\ni{n_\mathrm{i}}
\def\ne{n_\mathrm{e}}
\def\svie{\langle \sigma v \rangle_\mathrm{ie}}
\def\depsdpsi{\frac{\mathrm{d}\epsilon}{\mathrm{d}\psi}}
\def\dnidpsi{\frac{\mathrm{d}\ni}{\mathrm{d}\psi}}

\def\Ugo#1{\noindent{\color{red}#1}}

\title{Fast methods for tracking grain coagulation and ionization \\ III. Protostellar collapse with non-ideal MHD}
\titlerunning{}

\author{P. Marchand \inst{1,2}, U. Lebreuilly\inst{3}, M.-M. Mac Low \inst{2}, V. Guillet \inst{4,5}}

\institute{Institut de Recherche en Astrophysique et Planétologie, Université de Toulouse, UT3-PS, CNRS, CNES, 9 av. du Colonel Roche, 31028 Toulouse Cedex 4, France
\and Department of Astrophysics, American Museum of Natural History, 200 Central Park West, NY, NY, 10024, USA
\and AIM, CEA, CNRS, Universit\'e Paris-Saclay, Universit\'e Paris Diderot, Sorbonne Paris Cit\'e, 91191 Gif-sur-Yvette, France
\and Universit\'e Paris-Saclay, CNRS, Institut d’astrophysique spatiale, 91405, Orsay, France
\and Laboratoire Univers et Particules de Montpellier, Universit\'e de Montpellier, CNRS/IN2P3, CC 72, Place Eug\`ene Bataillon, 34095 Montpellier Cedex 5, France}

\authorrunning{P. Marchand et~al.}

\date{}

\abstract{Dust grains influence many aspects of star formation, including planet formation and the opacities for radiative transfer, chemistry, and the magnetic field via Ohmic, Hall, as well as ambipolar diffusion. The size distribution of the dust grains is the primary characteristic influencing all these aspects. Grain size increases by coagulation throughout the star formation process. In this work, we describe numerical simulations of protostellar collapse using methods described in earlier papers of this series. We compute the evolution of the grain size distribution from coagulation and the non-ideal magnetohydrodynamics effects self-consistently and at low numerical cost. We find that the coagulation efficiency is mostly affected by the time spent in high-density regions. Starting from sub-micron radii, grain sizes reach more than 100 $\mu$m in an inner protoplanetary disk that is only 1000 years old. We also show that the growth of grains significantly affects the resistivities, while also having an indirect effect on the dynamics and angular momentum of the disk.}


\maketitle

\section{Introduction}

Grains play a major role during star formation. First, they are the seeds of planet formation. While their characteristic size is sub-micron in the interstellar medium (ISM) \citep{mathis}, they grow by coagulation during the collapse and can reach sizes larger than 10 $\mu$m in the early stages of protostellar collapse \citep{2020A&A...643A..17G,2020A&A...641A..39S,2021ApJ...920L..35T,2022A&A...658A.191V,2022MNRAS.514.2145B}. Observations also suggest they reach these sizes (and possibly larger ones) in the envelopes of Class 0-I objects \citep{2009ApJ...696..841K, 2014A&A...567A..32M, 2019ApJ...885..106L,2019A&A...632A...5G}. Their growth then continues in protoplanetary disks until they eventually become planetesimals.
Variations in their size also significantly impact non-ideal magnetohydrodynamical (MHD) effects through their ionization and their chemical interactions with the gas, with direct feedback to the dynamics of the gas \citep{2016A&A...592A..18M,2016MNRAS.460.2050Z,2018MNRAS.473.4868Z,2020ApJ...900..180M,2020A&A...643A..17G}. Non-ideal MHD effects have been shown to be critical for the regulation of magnetic field and angular momentum during the protostellar collapse and the protoplanetary disk evolution \citep{mousse79,Machida_etal06,DuffinPudritz,MellonLi2009,LiKrasnopolskyShang,2015ApJ...801..117T,2016MNRAS.457.1037W,DA1,2018A&A...615A...5V,2020ApJ...900..180M,2021ApJ...917L..10L}. Grains are also the main source of opacity in protostellar environments, affecting the cooling of the gas and the observations made of those systems. Their high optical depth at densities of $\rho > 10^{-13}$ g cm$^{-3}$ leads to the formation of the first hydrostatic core \citep{Larson1969}.

However, numerical simulations usually do not account for grain coagulation self-consistently due to the great cost of computing a coagulation algorithm on the fly \citep[although new methods are being developed; see the recent work by][]{2021MNRAS.501.4298L}. The dust evolution used to be pre-processed or post-processed with no self-consistent feedback on the dynamics \citep{1991A&A...251..587R,2005A&A...434..971D,2016MNRAS.460.2050Z,2020ApJ...900..180M}. Recently, more and more studies include the growth of grains in their hydrodynamics simulations \citep{2021ApJ...920L..35T,2021MNRAS.507.2318V,2022A&A...658A.191V,2022MNRAS.514.2145B}. In \citet[][hereafter Paper I]{2021A&A...649A..50M}, we presented a simple and fast method to track coagulation in a self-consistent way that is particularly suited for modeling star formation. Here, we apply this method to non-ideal MHD protostellar collapse simulations. It is coupled with the second method presented in Paper I: a fast calculation of the ionization and grain charge to obtain non-ideal MHD resistivities. These 3D simulations are the first to include a self-consistent grain growth with direct feedback to the dynamics through the self-consistent calculation of MHD resistivities.

The paper is organized as follows. In Section \ref{sec:methods}, we describe the methods used in our study. Section \ref{sec:results} presents the results of our calculations, both analytical in Section \ref{Sec:analytical} and numerical in Section \ref{sec:numerical}. We compare our results to other works and discuss the caveats in Section \ref{sec:discussion}. We present our conclusions in Section \ref{sec:conclusions}.

\section{Methods}\label{sec:methods}

We performed non-ideal MHD simulations with the RAMSES code \citep{teyssier}. RAMSES is an Eulerian gas dynamics code with adaptive mesh refinement (AMR) and self-gravity. It includes a monofluid treatment of non-ideal MHD effects \citep{masson_nimhd,2018A&A...619A..37M}. We have implemented the methods presented in Paper I to calculate the coagulation and ionization of grains on-the-fly in a self-consistent manner.

\subsection{Grain coagulation and ionization}\label{SecMethodgrains}

\subsubsection{Coagulation}

In Paper I, we demonstrate that the coagulation process as described by the \citet{1916ZPhy...17..557S} equation is a one-dimensional (1D) process with certain types of coagulation kernels. Consequently, the size distribution of the coagulated grains depends only on the initial size distribution and a variable $\chi$ that encompasses the whole history of the physical conditions seen by the grains. At a given $\chi$ that is integrated along the path of the grains, the coagulated distribution is always the same, independently of the actual path taken. This method works for every coagulation kernel for which the dependence on the gas variables, such as density and temperature, can be separated from the grain properties such as size and mass. In Paper I and in the present paper, we use the turbulent kernel derived by \citet{2007A&A...466..413O} in the intermediate coupling regime, which is suited for star formation conditions. We show that in this case $\chi$ can be derived from integrating:
\begin{equation}\label{EqLagrangianchi}
    \mathrm{d}\chi = \nh ^{\frac{3}{4}} T ^{-\frac{1}{4}} \mathrm{d}t,
\end{equation}
where $\nh$ is the number density of the gas, $T$ its temperature, and $t$ the time.
In three-dimensional (3D) hydrodynamical simulations, only the knowledge of $\chi$ is needed to track the coagulation of grains.

Equation (\ref{EqLagrangianchi}) is a Lagrangian derivative of $\chi$ with respect to time along the path of the grain. We can transform it into a partial (Eulerian) derivative:
\begin{equation}\label{EqEulerianchi}
    \frac{\partial \chi}{\partial t} + \mathbf{u} \cdot \nabla \chi = n_\mathrm{H} ^{\frac{3}{4}} T ^{-\frac{1}{4}},
\end{equation}
where $\mathbf{u}$ is the velocity of the gas. We can combine Equation \ref{EqEulerianchi} with the mass conservation equation: 
\begin{equation}
    \frac{\partial \rho}{\partial t} +  \nabla \cdot [\rho \mathbf{u}]= 0,
\end{equation}
where $\rho$ is the gas mass density. This yields:
\begin{equation}
    \frac{\partial \rho \chi}{\partial t} + \nabla \cdot (\rho \chi \mathbf{u}) = \rho n_\mathrm{H} ^{\frac{3}{4}} T ^{-\frac{1}{4}},
\end{equation}
which means that the quantity $\rho \chi$ can be treated in an Eulerian framework as a passive scalar with a source term. We exploit this property and implement it as such in RAMSES. The value of $\chi$ is therefore calculated self-consistently in all cells at each time-step, as a mass-weighted average, ignoring any diffusion of dust through the gas. 
We used the Ishinisan code \citep{2021A&A...649A..50M} to pre-calculate a table containing the grain size distribution for a large number of $\chi$ values in a large relevant interval (between $\chi = 10^{13}$  and $\chi = 10^{19}$ in cgs units). When the size distribution is needed during the hydrodynamical simulation, it is interpolated from the table based on the value of $\chi$.

\begin{figure}
\begin{center}
\includegraphics[trim=3cm 2cm 2cm 2cm, width=0.45\textwidth]{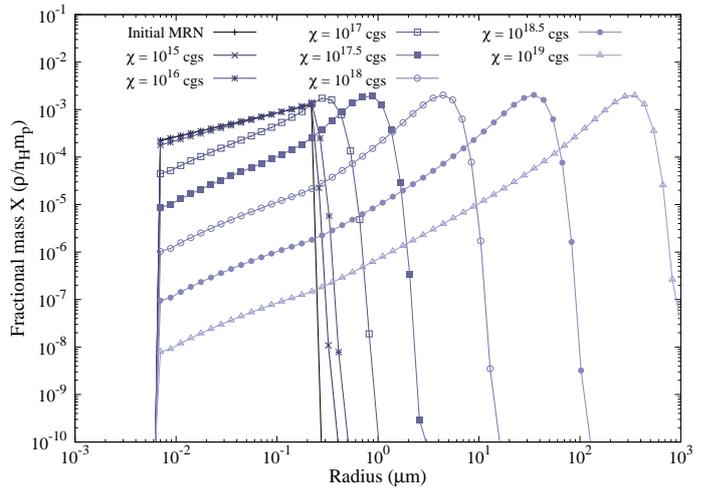}
  \caption{State of the grain size distribution for values of $\chi$, between $10^{15}$ cgs and $10^{19}$ cgs. The points represent the fractional abundance of the size-bin as function of the effective radius of the bin.}
  \label{FigDistribwithchi}
\end{center}
\end{figure}

Our initial distribution in this paper is a \citet[][]{mathis} distribution (i.e., MRN). The minimum and maximum radii are $a_\mathrm{min} = 5$~nm and $a_\mathrm{max} = 250$~nm, and the slope of the distribution is $-3.5$, so that the number density of grains, $n,$ follows the following variation:
\begin{equation}
    \frac{\mathrm{d}n}{\mathrm{da}} \propto a^{-3.5}.
\end{equation}
The total quantity of grains is determined by the dust-to-gas mass ratio that we assume to be 0.01. In this work, we sample the distribution with 60 bins of size logarithmically spaced between $5$ nm and $5000$ $\mu$m. 
In Figure \ref{FigDistribwithchi}, we present the coagulated MRN size distribution at various $\chi$. Below $\chi=10^{17}$ cgs, the shift in the size distribution is negligible. For higher values, the maximum size and the peak of the distribution are located at larger and larger radii, while the slope of the distribution remains similar. In all cases, the small grains are more abundant while the large grains hold more mass. The mode of the size distribution $a_\mathrm{max}$ is located near the largest relevant grain size.

As grains grow, they may experience fragmentation when the kinetic energy of the collision is high. Contrarily to what we derived in paper I, we find that fragmentation does not occur in early stages of the disk, but it is present, rather, only at very high densities $\rho > 10^{12}$  cm$^{-3}$ \citep{2023MNRAS.518.3326L} for very large grains with $a > 0.08$ cm. We demonstrate this result in Appendix \ref{AppFragmentation}. We therefore neglect fragmentation in this work.
We also do not account for the grain drift with respect to the gas in this work. We discuss the possible consequences in Section \ref{sec:largegrainsenvelope}. Drift is, however, compatible with our coagulation model, and we detail the method in Appendix \ref{AppDusttogas}. Other limitations are discussed in Section \ref{sec:kernels}.

\subsubsection{Ionization and resistivities}
\label{subsub:ion}
In Paper I, we also presented a fast method to calculate the ionization of the gas-grain mixture. For an arbitrary size distribution, we can calculate the average electric charge of each grain size, the number of ions, and the number of electrons, provided the cosmic-ray (CR) ionization rate, the density and temperature of the gas, the average atomic mass of ions, $\mu_\mathrm{i}$, and the sticking probability of electrons on grains, $s_\mathrm{e}$. Here, we assume $\mu_\mathrm{i} = 28$, which corresponds to the ion HCO$^+$, and $s_\mathrm{e} = 0.6$ as in \citet{2016A&A...592A..18M}. We also assume $\zeta = 5 \times 10^{-17}$ s$^{-1}$. The density and the temperature, are taken from the hydrodynamic simulation. The calculation is performed by the Newton-Raphson scheme described in Appendix A of Paper I.
The resistivities are computed using a similar method to \citet{2016A&A...592A..18M}, with one difference. For each grain size, they sum over the contributions of the whole charge distribution (between -1 to +1 in their case). Instead, we average the contributions using the mean electric charge. 
We explain the method in greater detail in Appendix \ref{AppIonization}.

\subsection{Star formation simulations}

\subsubsection{Model}

We performed four numerical simulations using the RAMSES code. We solved the following MHD equations:
\begin{align}
  &\frac{\partial \rho}{\partial t} + \nabla \cdot \left[\rho \mathbf{u}\right]  = 0, \label{eqmass}\\
  &\frac{\partial \rho \mathbf{u}}{\partial t} + \nabla \cdot \left[\rho \mathbf{u}\mathbf{u} + \left(P+\frac{B^2}{2}\right) \mathbb{I} - \mathbf{BB} \right]  = -\rho \nabla \Phi, \label{eqmom} \\ 
  &\frac{\partial \mathbf{B}}{\partial t} - \nabla \times \left[\mathbf{u} \times \mathbf{B} - \eta_\Omega \mathbf{J} - \eta_\mathrm{H} \frac{\mathbf{J}\times \mathbf{B}}{B}+ \eta_\mathrm{AD}\frac    {\left(\mathbf{J} \times \mathbf{B}\right) \times \mathbf{B}}{B^2} \right]  = 0,   \label{eqinduc} \\
  &\nabla \cdot \mathbf{B}=0,\label{nablab} 
\end{align}
where $\mathbf{u}$ is the velocity of the gas, $P$ its pressure, $\mathbf{B}$ the magnetic field, $\mathbf{J} = \nabla \times \mathbf{B}$ the current, $\mathbb{I}$ the identity matrix, $\Phi$ the gravitational potential, and $\eta_\Omega$, $\eta_\mathrm{H}$, and $\eta_\mathrm{AD}$ are the ohmic, Hall, and ambipolar resistivities, respectively.
The temperature evolution is prescribed by the barotropic equation:
\begin{equation}
T=T_0\left(1+\left[ \frac{\rho}{10^{-13}~\mathrm{g}~\mathrm{cm}^{-3}}\right]^{\gamma-1}\right),
\end{equation}
with $T_0=10$ K and $\gamma=5/3$ as the adiabatic index. The initial condition is a sphere of gas of 1 M$_\odot$. The radius of the sphere is controlled by the thermal over gravitational energy ratio $\alpha$. We chose $\alpha=0.3$, which sets a radius of $R=2946$ au. The density distribution has an $m=2$ azimuthal perturbation of
\begin{equation}
\rho(\varphi) = \rho_0 (1+\delta_\rho \sin\varphi),
\end{equation}
where $\rho_0$ is the density of a uniform sphere of same mass and radius, and $\varphi$ the azimuthal angle. We choose $\delta_\rho=0.05$. The computational domain outside of the sphere is filled with gas of density $\rho_0/100$. The sphere undergoes a solid rotation, with a ratio of rotational to gravitational energy of $\beta=0.02$. The magnetic field is initially uniform and parallel to the rotation axis. It is defined using the mass-to-flux ratio over the critical value \citep{MouschoviasSpitzer1976}:
\begin{equation}
  \mu_\mathrm{B} = \frac{M/\Phi_\mathrm{B}}{\left(M/\Phi_\mathrm{B}\right)_\mathrm{crit}},
\end{equation}
\noindent with
\begin{equation}
  \left(\frac{M}{\Phi_\mathrm{B}}\right)_\mathrm{crit}=\frac{0.53}{3\pi}\sqrt{\frac{5}{G}}.
\end{equation}

Observations show that dense cores are slightly super-critical \citep{1999ApJ...520..706C}, although recent numerical simulations indicate that observations may overestimate the actual strength of the magnetic field due to projection effects \citep{2020ApJ...893...73K}. We chose $\mu_\mathrm{B}=5$ as a fiducial value.
Those parameters are used in our reference case C-3. We changed the value of $\alpha$ to 0.4, and the initial mass, $M,$ to $5$ M$_\odot$ for two other cases referred to as C-4 and C-3-M5, respectively, to investigate the influence of the collapse time on the grain coagulation. The final run, named NC-3, is similar to C-3 without coagulation.
The grain distribution and ionization evolve as described in Sections \ref{SecMethodgrains} and \ref{subsub:ion}, and in Paper I. The four cases are summarized in Table \ref{TableSetups}.

\begin{table*}
  \caption{Parameters of the protostellar collapse simulations: name of the simulation, thermal over gravitational energy ratio $\alpha$, radius R, mass M, initial density, $\rho_0$, and initial magnetic field B of the initial cloud, formation time of the first hydrostatic core, $t_\mathrm{fc}$, and use of coagulation.}
\label{TableSetups}
\centering
\begin{tabular}{llllllll}
\hline\hline
  Name   &  $\alpha$  & R (au)      & M (M$_\odot$) & $\rho_0$ (g cm$^{-3}$) &B ($\mu$G) & $t_\mathrm{fc}$ (kyr) & Coagulation\\
\hline
 C-3   &  $0.3$     & $2946$      & $1$  & $5.49 \times 10^{-18}$ & 133 & $\sim 30$ & Yes\\
 C-3-M5   &  $0.3$     & $14738$      & $5$ & $2.19 \times 10^{-19}$  & 27 & $\sim 150$ & Yes\\
 C-4   &  $0.4$     & $3930$      & $1$ & $2.31 \times 10^{-18}$  & 75 & $\sim 47$& Yes\\
 NC-3    &  $0.3$     & $2946$      & $1$ & $5.49 \times 10^{-18}$  & 133 & $\sim 30$ & No\\
\hline
\end{tabular} 
\end{table*}

\subsubsection{Grid and algorithm}

The simulation box is a cube that is four times as large as the radius of the sphere, with periodic boundary conditions. The initial grid is composed of $32^3$ cells (level 5 of AMR) and is refined to ensure at least ten points per Jeans length, strongly satisfying the \citet{1997ApJ...489L.179T} criterion. The maximum AMR refinement level is 13 for C-3 and NC-3 (resolution of 1.4 au), 14 for C-4 (resolution of 0.96 au), and 16 for C-3-M5 (resolution of 0.90 au). 

Simulations were performed with a 3D unsplit slope limiter to avoid overshooting of the magnetic field at shock boundaries, while keeping the second order convergence for the Hall effect. We used the HLLD Riemann solver \citep{2005JCoPh.208..315M} for non-magnetic variables and the 2D HLL Riemann solver for the magnetic field and the Hall effect \citep{2012JCoPh.231.7476B,2018A&A...619A..37M}. 
The Poisson equation is solved using the multigrid method of \citet{2011JCoPh.230.4756G} in our periodic domain.

\section{Application to star formation} \label{sec:results}

In this section, we present the results of our calculations. We first applied our coagulation method to an analytical one-zone model in Section \ref{Sec:analytical}, then in 3D MHD simulations in Section \ref{sec:numerical}.

\subsection{Analytical collapse} \label{Sec:analytical}

During spherical protostellar collapse, the time evolution of the contraction ratio of a gas cloud compared to its original radius $x=R(t)/R_0$ can be described as  \citep{2005A&A...436..933F}:
\begin{equation}
  \frac{\mathrm{d}x}{\mathrm{d}t} = - \frac{\pi}{2\tau_\mathrm{ff}}\sqrt{\frac{1}{x}-1},
\end{equation}
where $\tau_\mathrm{ff}$ is the free-fall time, given by:
\begin{equation}
  \tau_\mathrm{ff} = \sqrt{\frac{3\pi}{32G\rho_0}},
\end{equation}
where $\rho_0$ is the initial density. \citet{2020A&A...643A..17G} showed that assuming a uniform compression of the gas nicely reproduces the isothermal phase of the collapse, particularly when comparing the dust size distribution at the same gas density. In this case, the density of mass scales as $\rho(t)=\rho_0(R_0/R[t])^3$. The gas density then evolves as:
\begin{equation}
  \frac{1}{n_\mathrm{H}}\frac{\mathrm{d}n_\mathrm{H}}{\mathrm{d}t} =  \frac{1}{\rho}\frac{\mathrm{d}\rho}{\mathrm{d}t} = -\frac{3}{x}\frac{\mathrm{d}x}{\mathrm{d}t}.
\end{equation}

\begin{figure}
\begin{center}
\includegraphics[trim=1cm 1cm 2cm 0cm, width=0.45\textwidth]{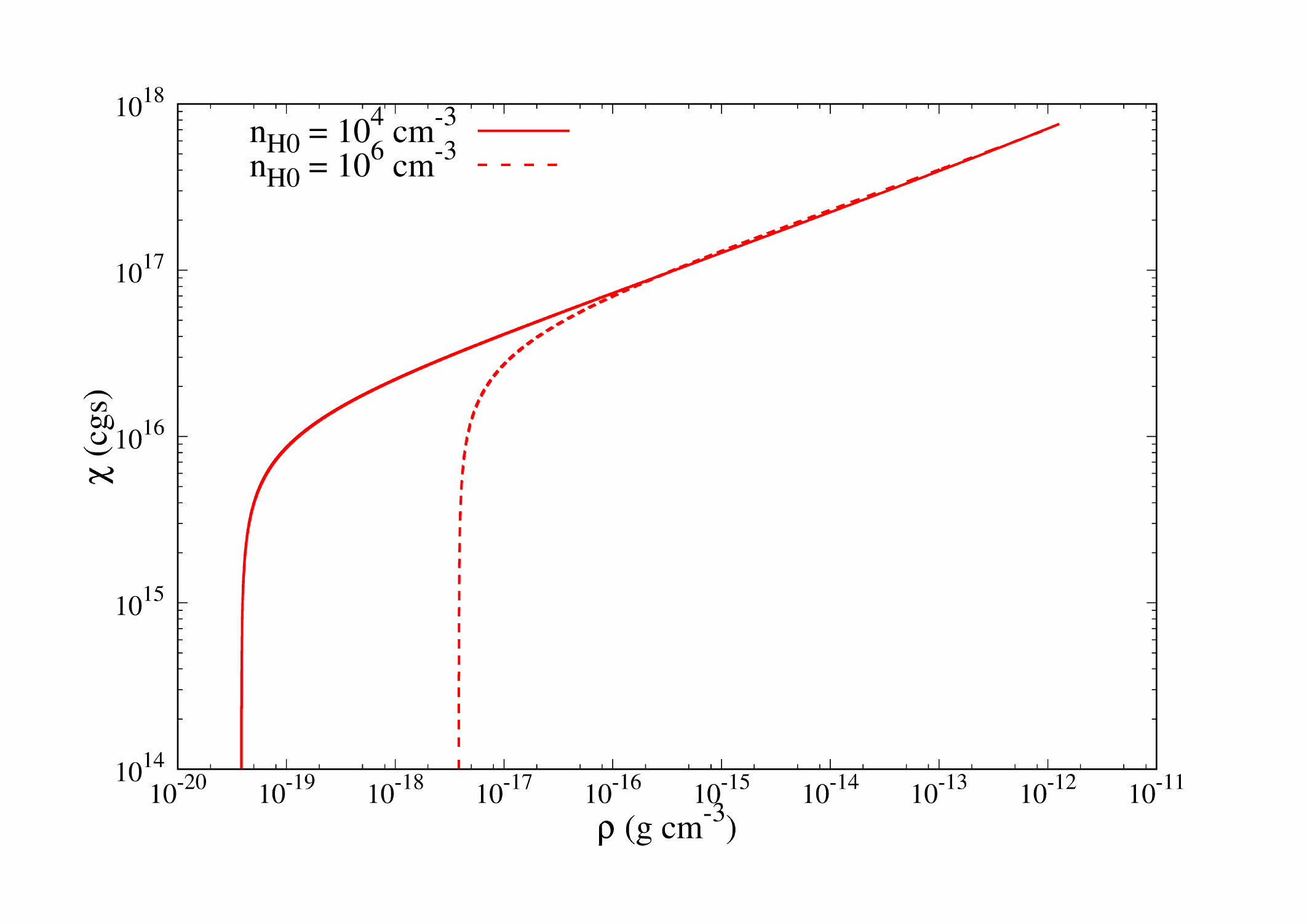}
  \caption{Evolution of the coagulation variable, $\chi,$ with increasing density, $\nh$, during the isothermal collapse. The solid line represents an initial density of $\rho_0 = 3.8 \times 10^{-20}~\gcc$ is and the dashed line represents $\rho_0 = 3.8 \times 10^{-18}~\gcc$. }
  \label{FigChivsnh}
\end{center}
\end{figure}

We used a second-order Runge-Kutta scheme to numerically integrate this equation from the beginning of the collapse at $\rho = \rho_0$ until the formation of the first Larson core at $\rho = 10^{-13}~\gcc$. The evolution of $\chi$ with $\rho$ is plotted in Figure~\ref{FigChivsnh}, assuming $T=10$ K. The solid and dashed lines represent different initial densities, $\rho_0 = 3.8\times10^{-20}~\gcc$ and $\rho_0 = 3.8\times10^{-18}~\gcc$, respectively. At densities $\nh > 10^{-16}~\gcc$, the value of $\chi$ is independent from $\rho_0$ and increases as $\chi \propto \rho^{1/4}$. This evolution is expected as $\chi \sim \rho^{3/4} t$ and $t \sim \tau_\mathrm{ff} \sim \rho^{-1/2}$. At $\rho = 10^{-13}~\gcc$, the coagulation variable reaches $\chi \approx 5.6 \times 10^{17}$ cgs, which corresponds to a peak of the size distribution of $\sim 10$ $\mu$m, indicating a significant grain growth. That is consistent with the results of \citet{2020A&A...643A..17G}, who use the same collapse model, but solve coagulation on the fly.

\subsection{Numerical collapse} \label{sec:numerical}

\begin{figure}
\begin{center}
\includegraphics[trim=6cm 2cm 5cm 2.5cm, width=0.5\textwidth]{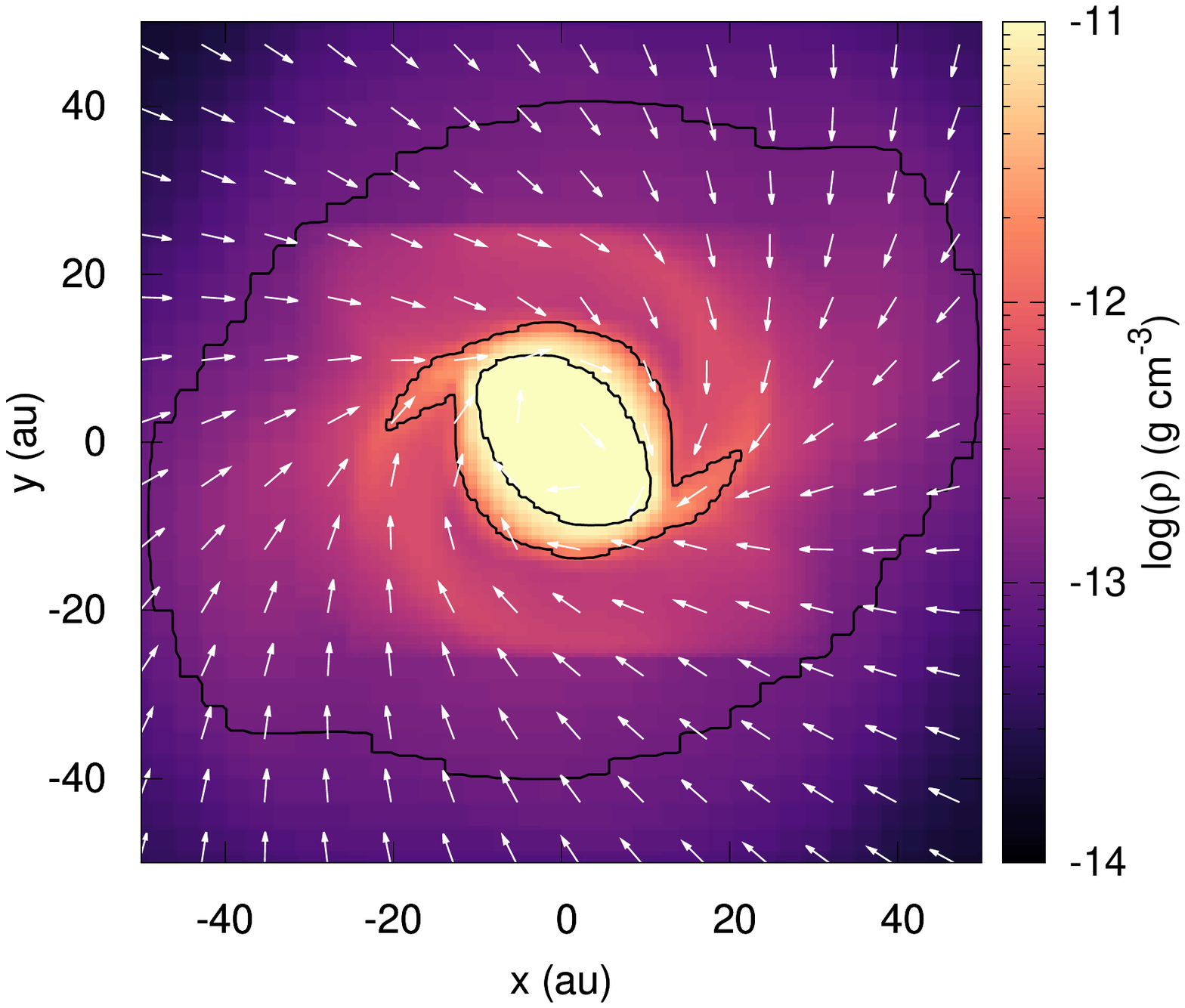}
\includegraphics[trim=6cm 2cm 5cm 3.5cm, width=0.5\textwidth]{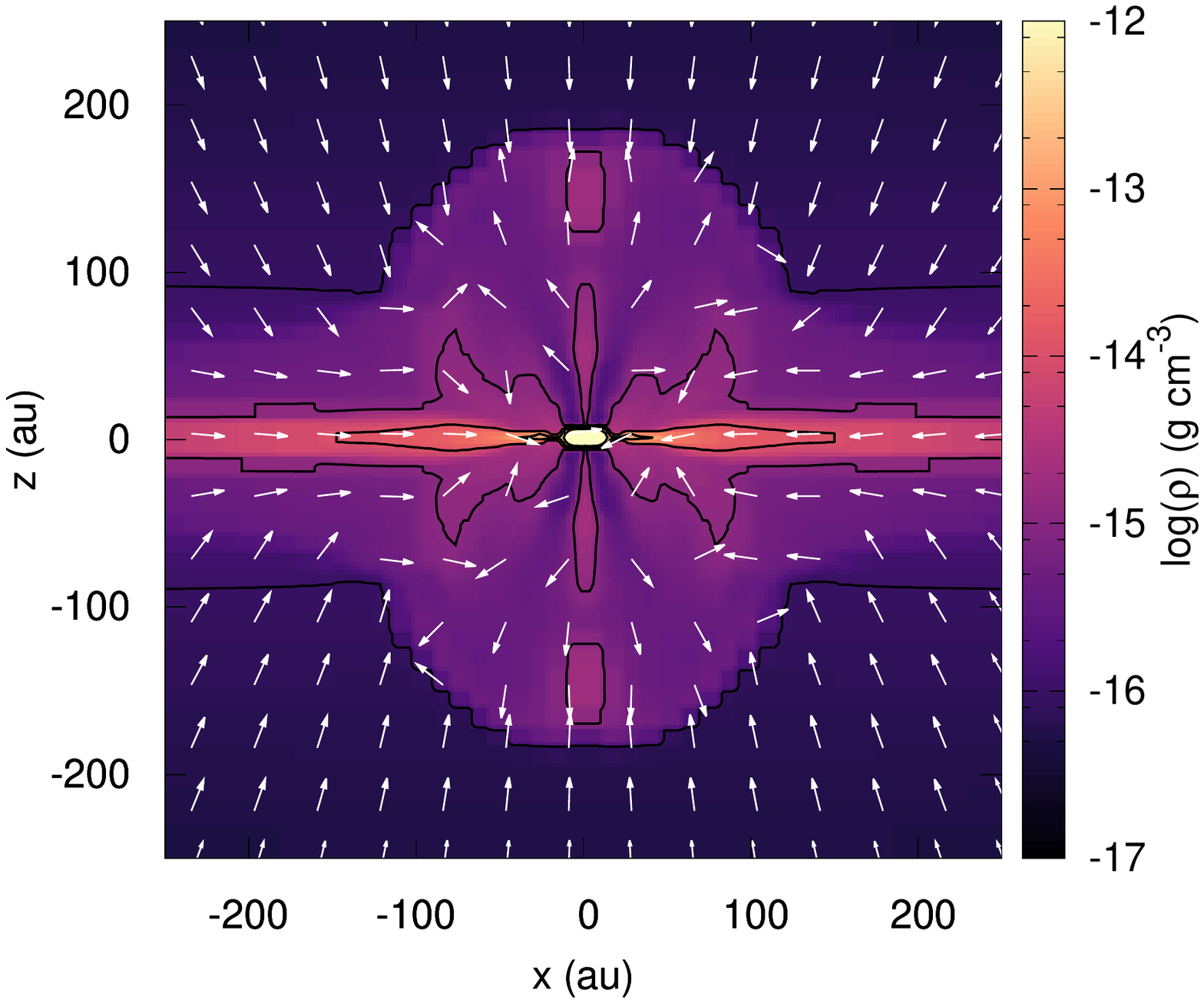}
  \caption{Density slices of the C-3 simulation, with grain coagulation. Top panel is a face-on slice of the plane z=0, and the bottom panel an edge-on slice of the plane y=0. White arrows represent the direction of the gas velocity. The snapshot is taken at the final time-step, 1000 years after the formation of the first Larson core.}
  \label{FigMapscoag}
\end{center}
\end{figure}

The numerical simulations were run as described in Section \ref{sec:methods} until 1000 years after the density reaches $10^{-13}$ g cm$^{-3}$, the formation of the first hydrostatic core at $\sim 30$~kyr (Tab.~\ref{TableSetups}). In reference simulation C-3, a small circumstellar disk with a radius of $\approx$ 20 au forms and a disk wind is launched by magneto-centrifugal acceleration \citep{1982MNRAS.199..883B}. Figure \ref{FigMapscoag} shows face-on and edge-on slices of density at the final time-step of the simulation, with arrows indicating the gas velocity.

\subsubsection{Grain growth}

\begin{figure}
\begin{center}
\includegraphics[trim=2cm 1cm 2cm 0.5cm, width=0.45\textwidth]{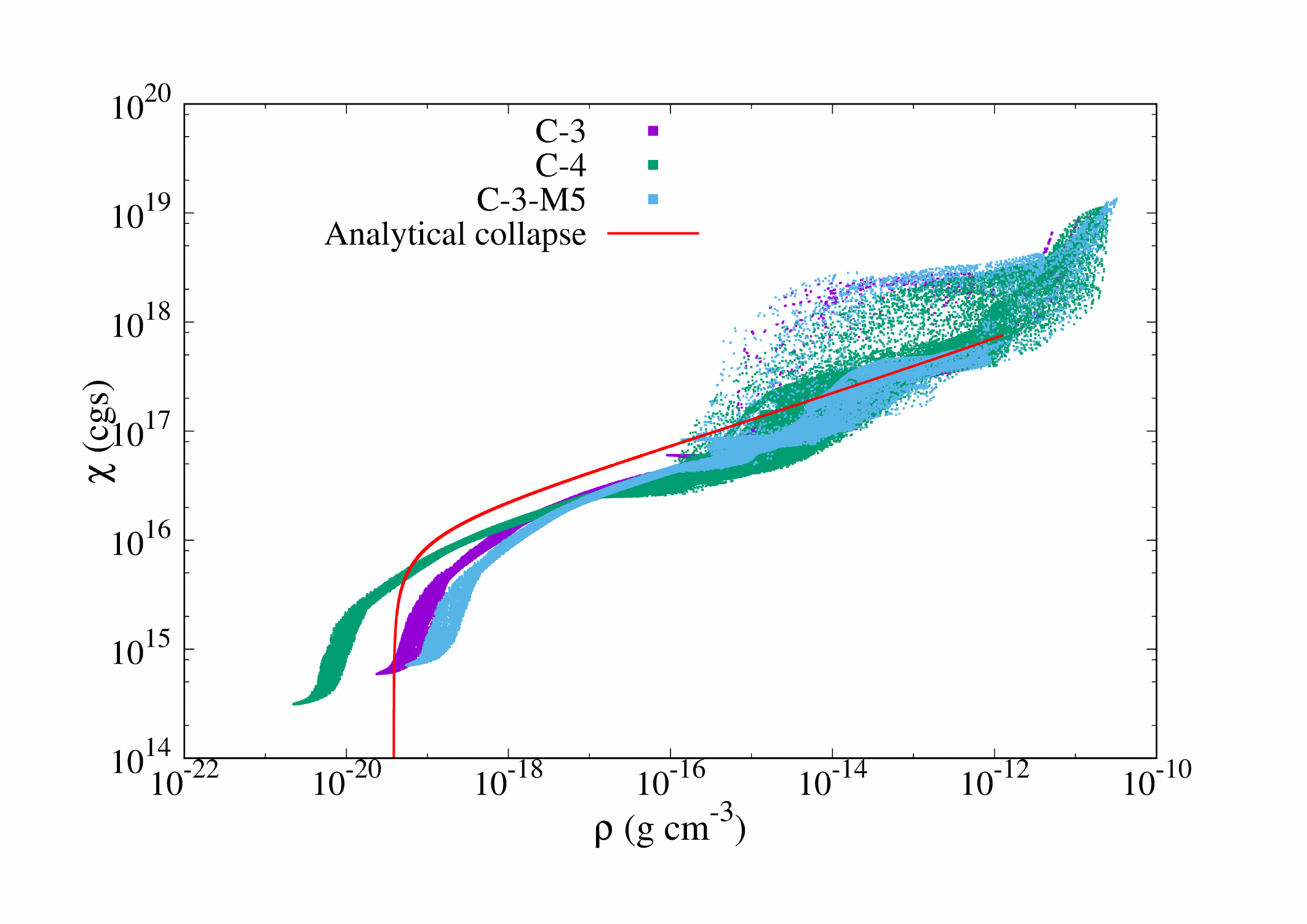}
  \caption{Evolution of the coagulation variable $\chi$ with increasing density $\nh$ in the numerical collapse models C-3 (purple), C-3-M5 (light blue), and C-4 (green). Each point corresponds to a simulation cell at the final time-step. The red line is the analytical collapse solution for $\rho_0=3.8 \times 10^{-20}$ g cm$^{-3}$.}
  \label{FigChivsnhSimu}
\end{center}
\end{figure}

We show in Figure \ref{FigChivsnhSimu} the value of $\chi$ as a function of the density in simulation cells at the final time-step, for runs C-3, C-4, and C-3-M5. The increase is quasi unidimensional in the isothermally collapsing envelope for $\rho < 10^{15}$ g cm$^{-3}$. Beyond this density, there is a large spread of $\chi$ values of over an order of magnitude. This spread most likely occurs in gas falling in the pseudo-disk, then the disk and the first Larson core, or the outflow, over different timescales, spending unequal times in a given density range. The overall trend agrees well with the analytical calculation. 

There is no significant difference in the $\chi$ values, and thus dust size distributions, between the three runs, despite run C-4 needing 50$\%$ more time to collapse to the first Larson core stage, and C-3-M5 needing $400\%$ more time. Coagulation happening in the isothermally collapsing envelope is therefore hardly impacted by the initial conditions, as growth accelerates with increasing density.

\begin{figure}
\begin{center}
\includegraphics[trim=2cm 1cm 2cm 1cm, width=0.45\textwidth]{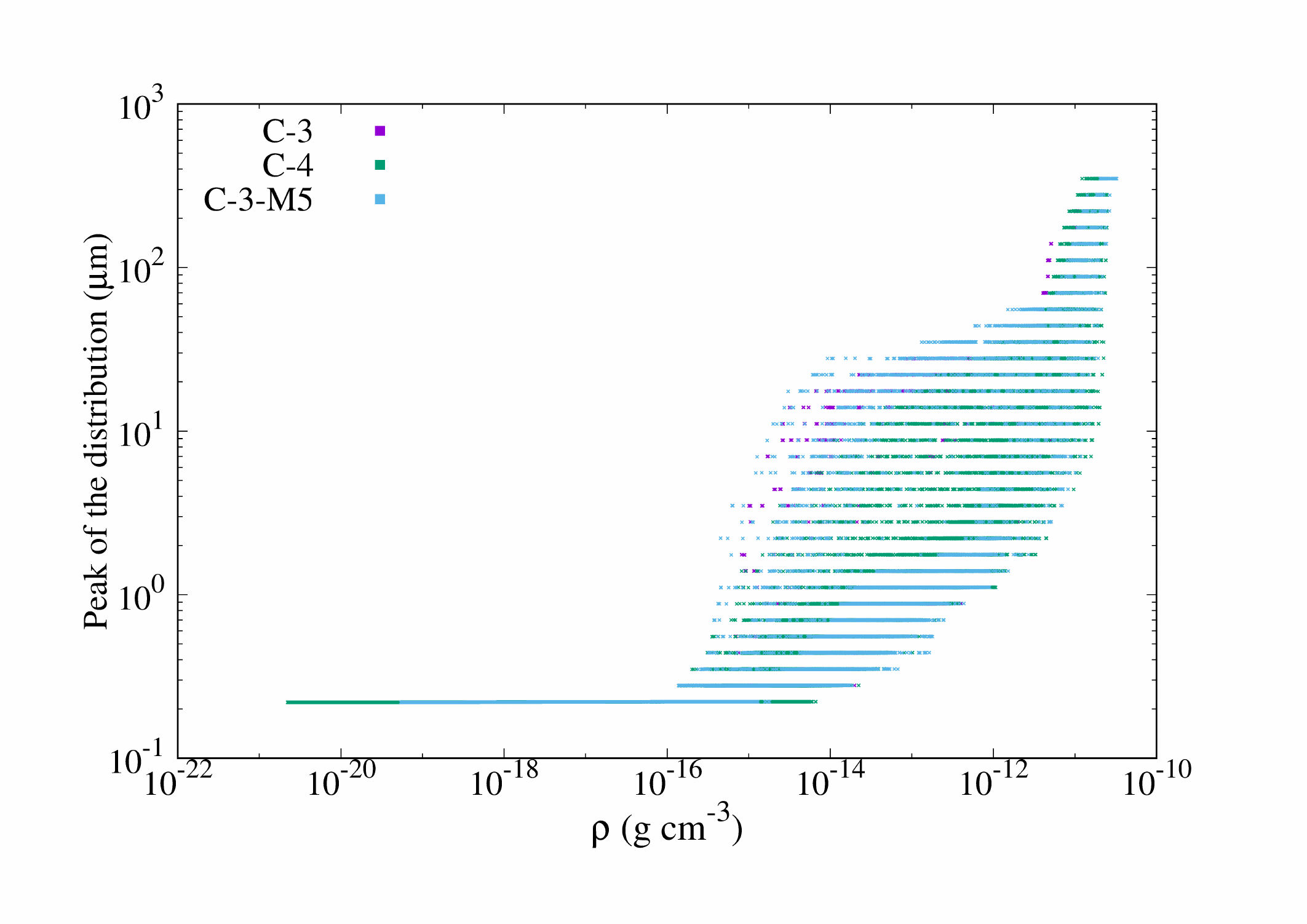}
  \caption{Mode of the coagulated grain size distribution as a function of density in simulations (purple), C-3-M5 (light blue), and C-4 (green). The discrete values of the sizes are due to the binning of the size distribution.}
  \label{FigAmaxvsnh}
\end{center}
\end{figure}

Figure \ref{FigAmaxvsnh} shows the mode of the size distribution as a function of density, corresponding to the distribution of $\chi$ shown in Figure \ref{FigChivsnhSimu}. Three regions are clearly demarcated, the first being the envelope, in which grain coagulation is not efficient enough to form large grains ($\rho < 10^{-16}$ g cm$^{-3}$). The second comprises the pseudo-disk and the early protoplanetary disk, where grains grow by a factor 100 from sub-micron sizes to several tens of $\mu$m. The third region is the first Larson core, which has even larger grains that reach 400 $\mu$m within a mere $10^3$~yr years after its formation. That value is in line with similar recent studies \citep{2022MNRAS.515.2072K,2023MNRAS.518.3326L}. There is little difference between runs C-3, C-4, and C-3-M5, confirming that coagulation in the envelope does not impact large grains, as found by previous studies \citep[for example][]{2022ApJ...940..188S}. We discuss observations of large grains in the envelope in Section~\ref{sec:largegrainsenvelope}.

\begin{figure}
\begin{center}
\includegraphics[trim=6cm 2cm 5cm 2.5cm, width=0.5\textwidth]{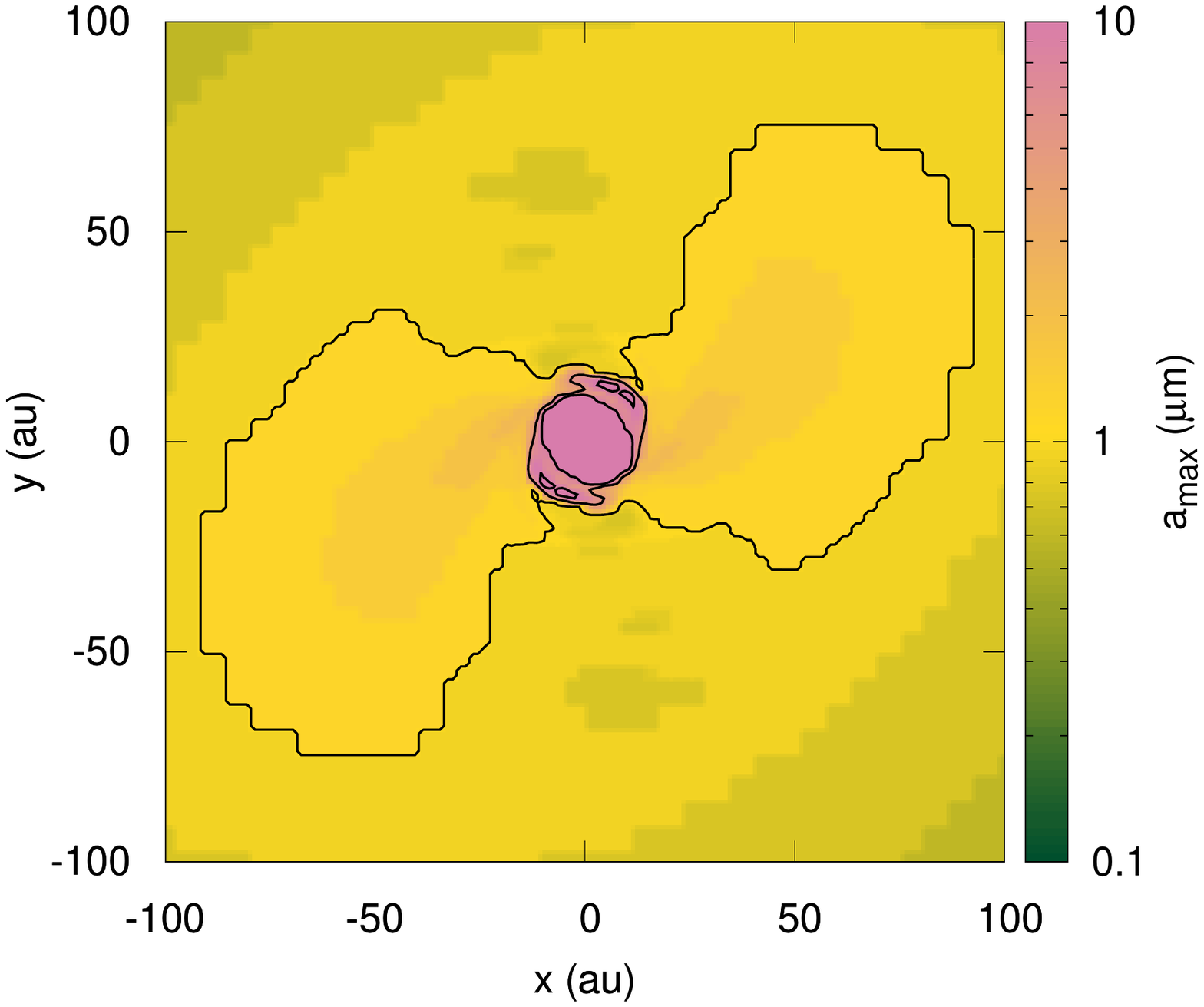}
\includegraphics[trim=6cm 2cm 5cm 3.5cm, width=0.5\textwidth]{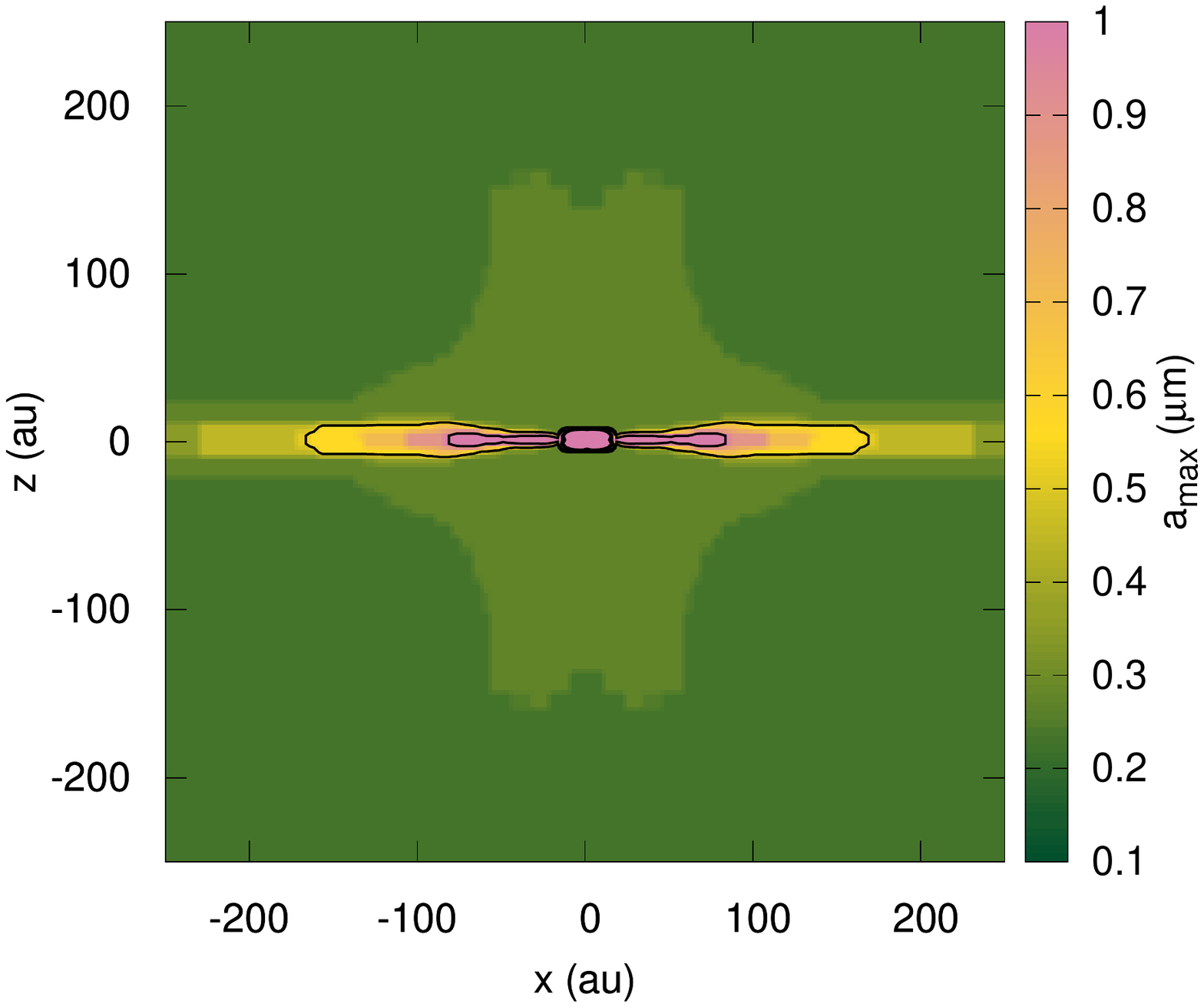}
  \caption{Slices of the C-3 simulation with coagulation showing the mode of the grain size distribution $a_\mathrm{max}$, 1000 years after the first core formation. The color scale is different for each panel to yield a better contrast. Top panel: Black contours indicate $a_\mathrm{max}=1$, 5 and 10 $\mu$m. Bottom panel:  $a_\mathrm{max}=0.5$ and 1 $\mu$m.}
  \label{FigMapamax}
\end{center}
\end{figure}

The spatial distribution of those grain sizes for run C-3 is displayed in Figure \ref{FigMapamax}. Size distributions shifting significantly from the initial MRN distribution are indeed confined to the mid-plane in the disk and pseudo-disk. The bottom panel also shows moderately larger grains in the outflow, as they only traveled through the upper layers of the pseudo-disk before being ejected \citep{2020ApJ...900..180M}. If they had been in the mid-plane of the disk, they would have grown much more, as coagulation is irreversible in our model. The upper panel shows that grains reach a radius larger than 1 $\mu$m in the outskirts of the disk, within 100 au of the center. Growth then occurs rapidly as density increases in the inner 15 au, which is shown by the almost overlapping contours delimiting $a_\mathrm{max}=5$ $\mu$m and $a_\mathrm{max}=10$ $\mu$m.

\begin{figure}[ht!]
\begin{center}
\includegraphics[trim=3cm 2cm 3cm 1.5cm,  width=0.45\textwidth]{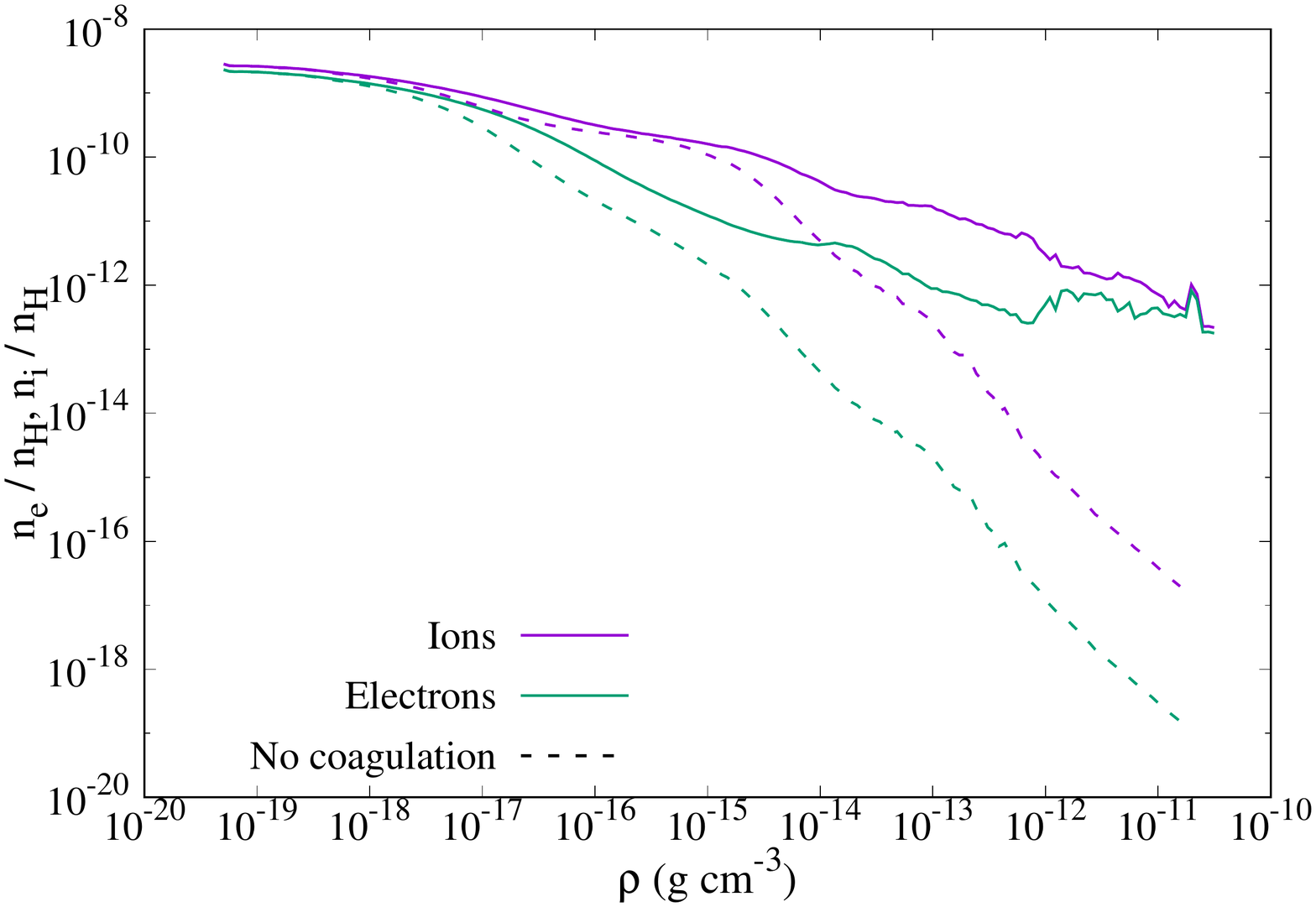}
\includegraphics[trim=3cm 1cm 3cm 0cm, width=0.45\textwidth]{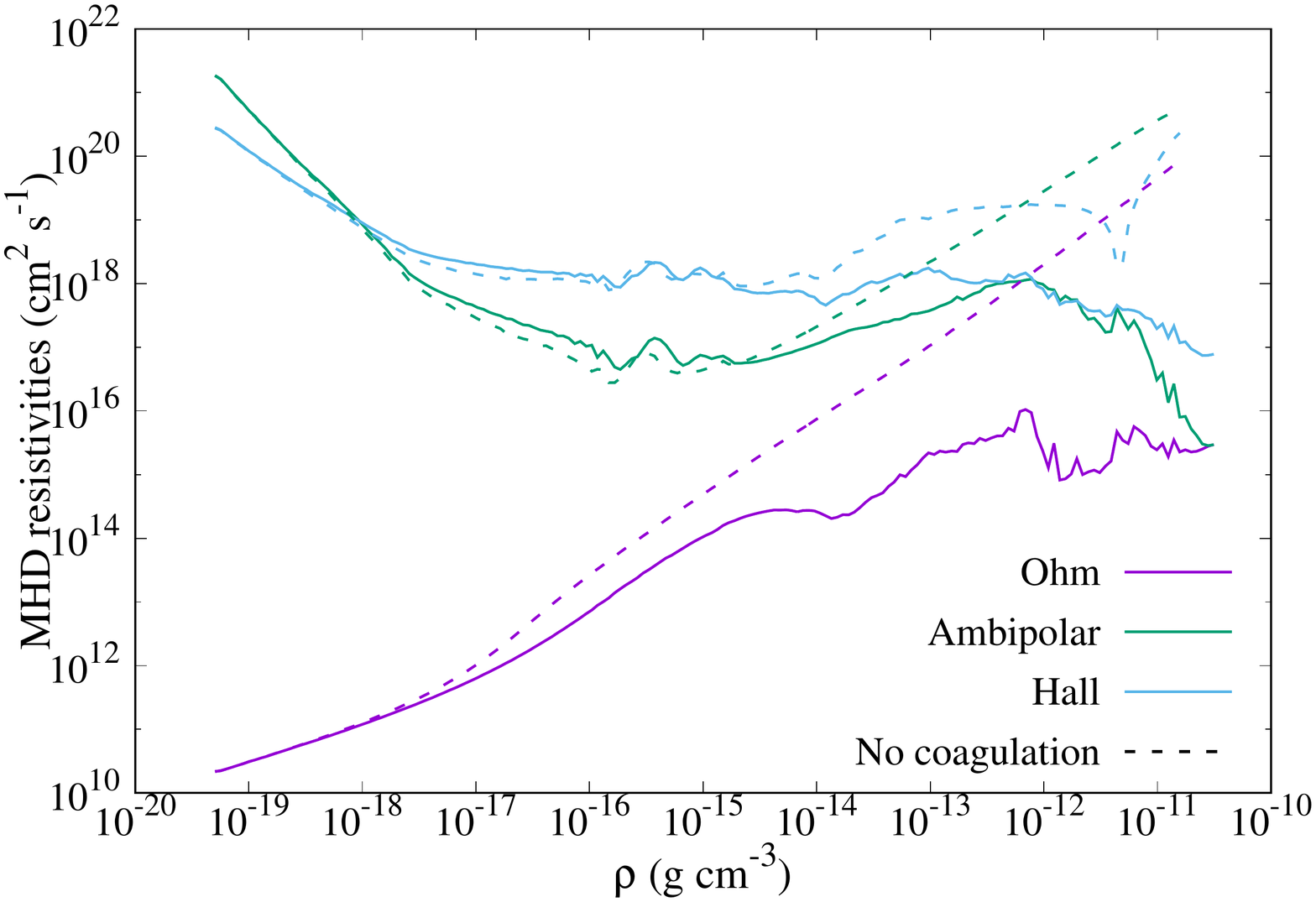}
\includegraphics[trim=3cm 1cm 3cm 1cm, width=0.45\textwidth]{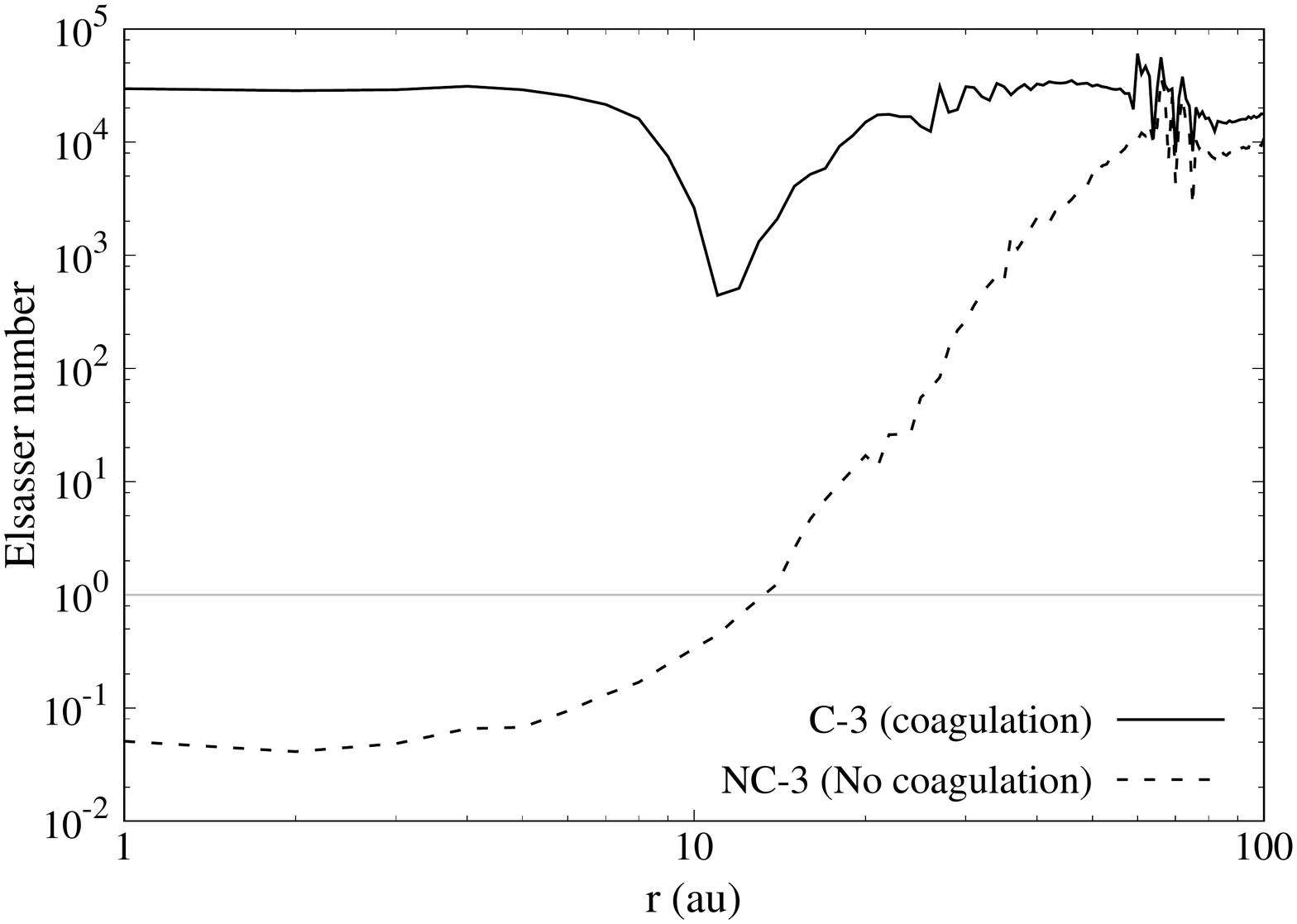}
  \caption{
  Non-ideal MHD effects in the C-3 and NC-3 simulations. Top panel: Abundances of ions (purple) and electrons (green) as a function of density in models with coagulation (C-3; solid lines) or without it (NC-3; dashed lines). Middle panel: Volume-averaged  Ohmic (purple), ambipolar (green), and Hall (blue) resistivities for the same models. Bottom panel: Average ambipolar Elsasser number $\mathrm{Am}$ in the mid-plane as a function of radius. The thin grey line represents $\mathrm{Am}=1$.}
  \label{FigAveresist}
\end{center}
\end{figure}

\subsubsection{Resistivities and gas dynamics}

We describe here the impact of grain coagulation on non-ideal MHD resistivities and their macroscopic effects on gas dynamics. Previous studies emulated the coagulation of grains by removing the very small grains ($a<0.1$ $\mu$m) and redistributing their mass to the larger end of the distribution \citep{2016MNRAS.460.2050Z,2018MNRAS.473.4868Z,2020ApJ...900..180M}. This method leads to an increase in resistivities, in particular the ambipolar resistivity, resulting in weaker coupling between the magnetic field and the gas, hence weaker magnetic braking and larger, more unstable disks. However, what we observe here is the exact opposite behavior.

\begin{figure}
\begin{center}
\includegraphics[trim=6cm 2cm 5.5cm 2.5cm, width=0.5\textwidth]{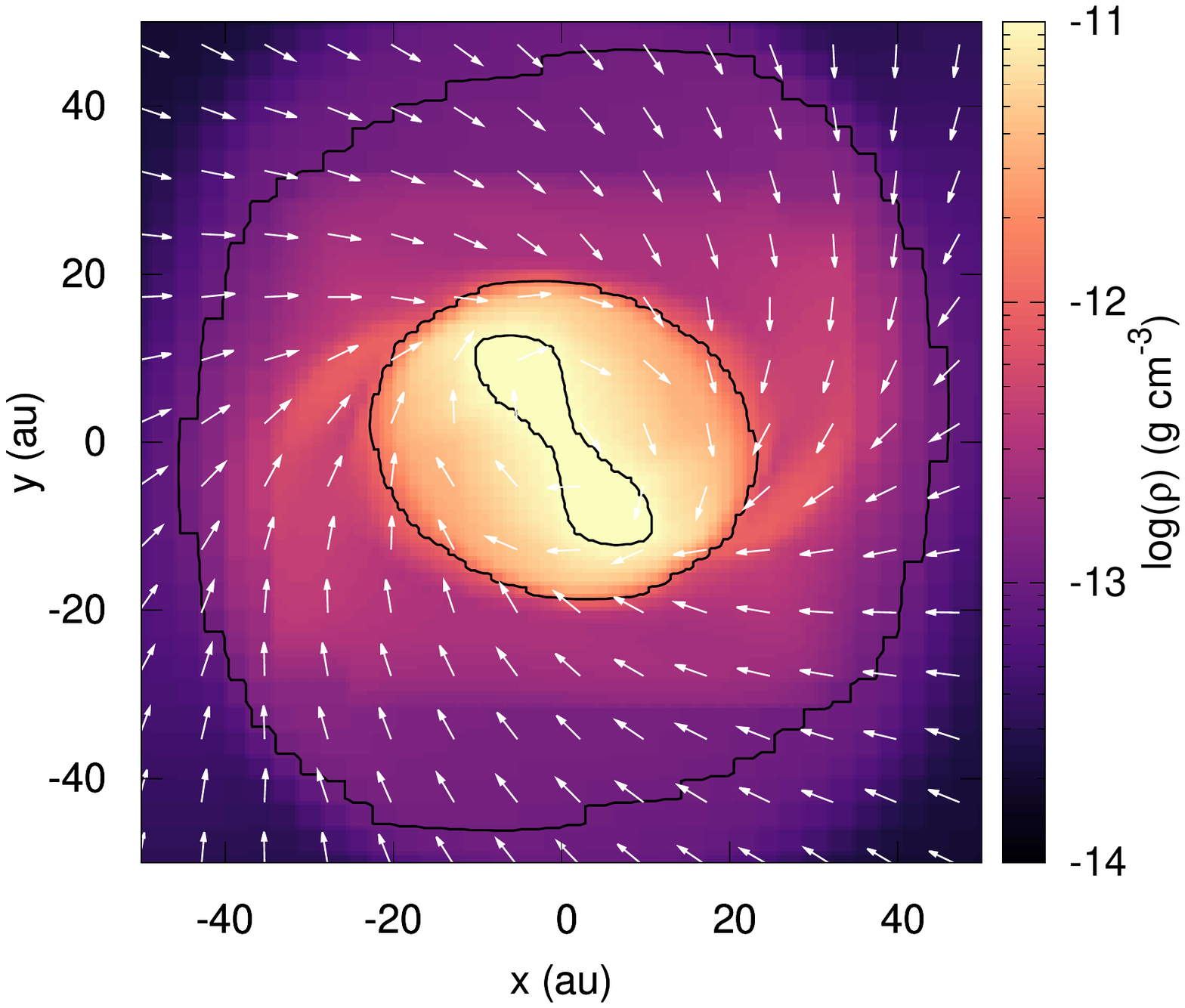}
\includegraphics[trim=6cm 2cm 5.5cm 3.5cm, width=0.5\textwidth]{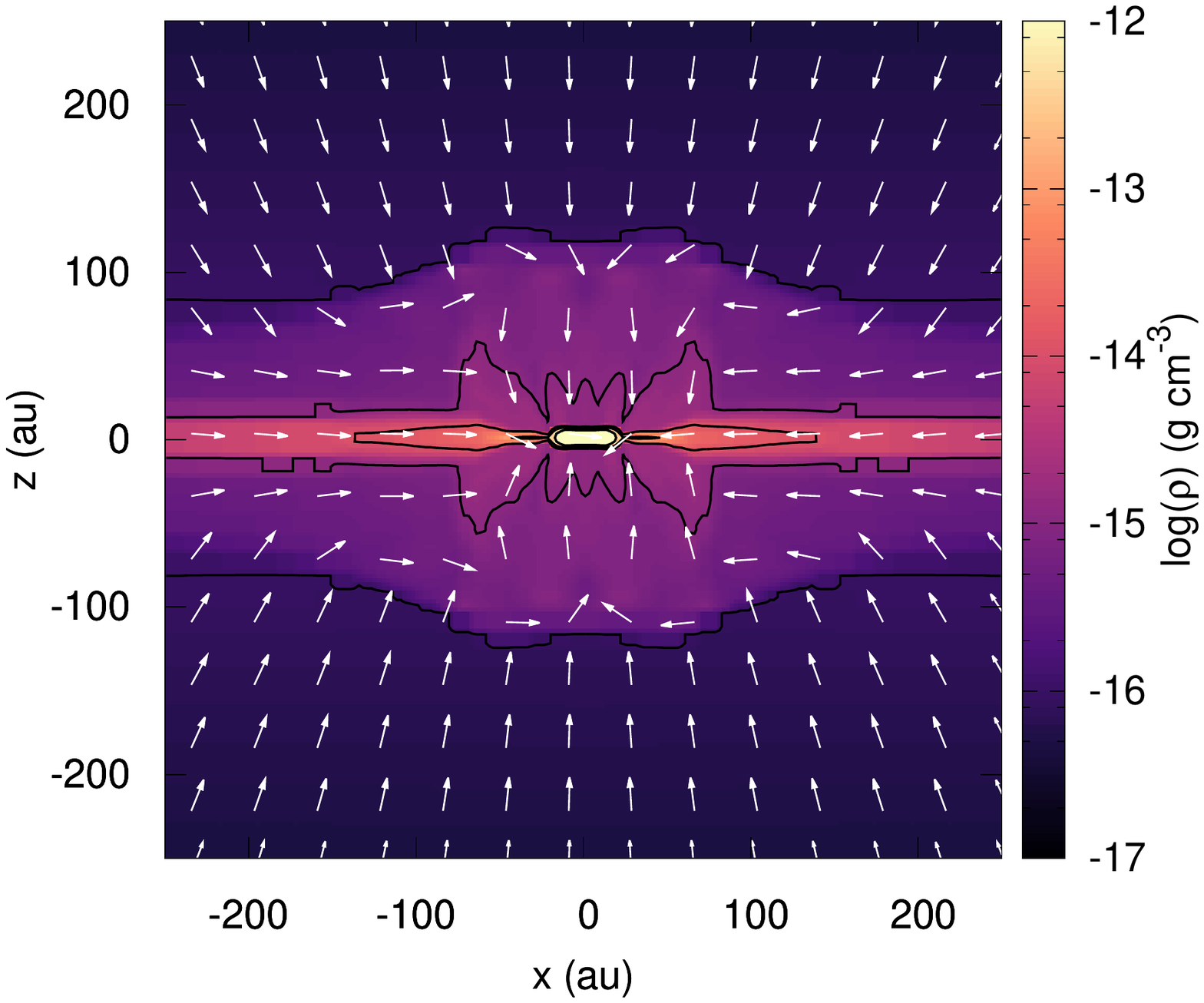}
  \caption{Density slices of the NC-3 simulation, without grain coagulation. Top panel is a face-on slice of the plane z=0, and the bottom panel an edge-on slice of the plane y=0. White arrows represent the direction of the gas velocity. The snapshot is taken at the final time-step, 1000 years after the formation of the first Larson core.}
  \label{FigMapsnocoag}
\end{center}
\end{figure}

The middle panel of figure \ref{FigAveresist} presents the volume-weighted average resistivities as a function of density, at the final time-step, for simulations C-3 and NC-3. A non-evolving size distribution produces resistivities relatively similar in the envelope, but two to four orders of magnitude larger at disk densities, particularly the Ohmic and ambipolar resistivities. 
Consequently, the magnetic braking is weakened, and the gas retains more angular momentum without the grain coagulation, forming a larger disk, as shown in Figure~\ref{FigMapsnocoag}. 

\begin{figure}
\begin{center}
\includegraphics[trim=3cm 1cm 3cm 0cm, width=0.45\textwidth]{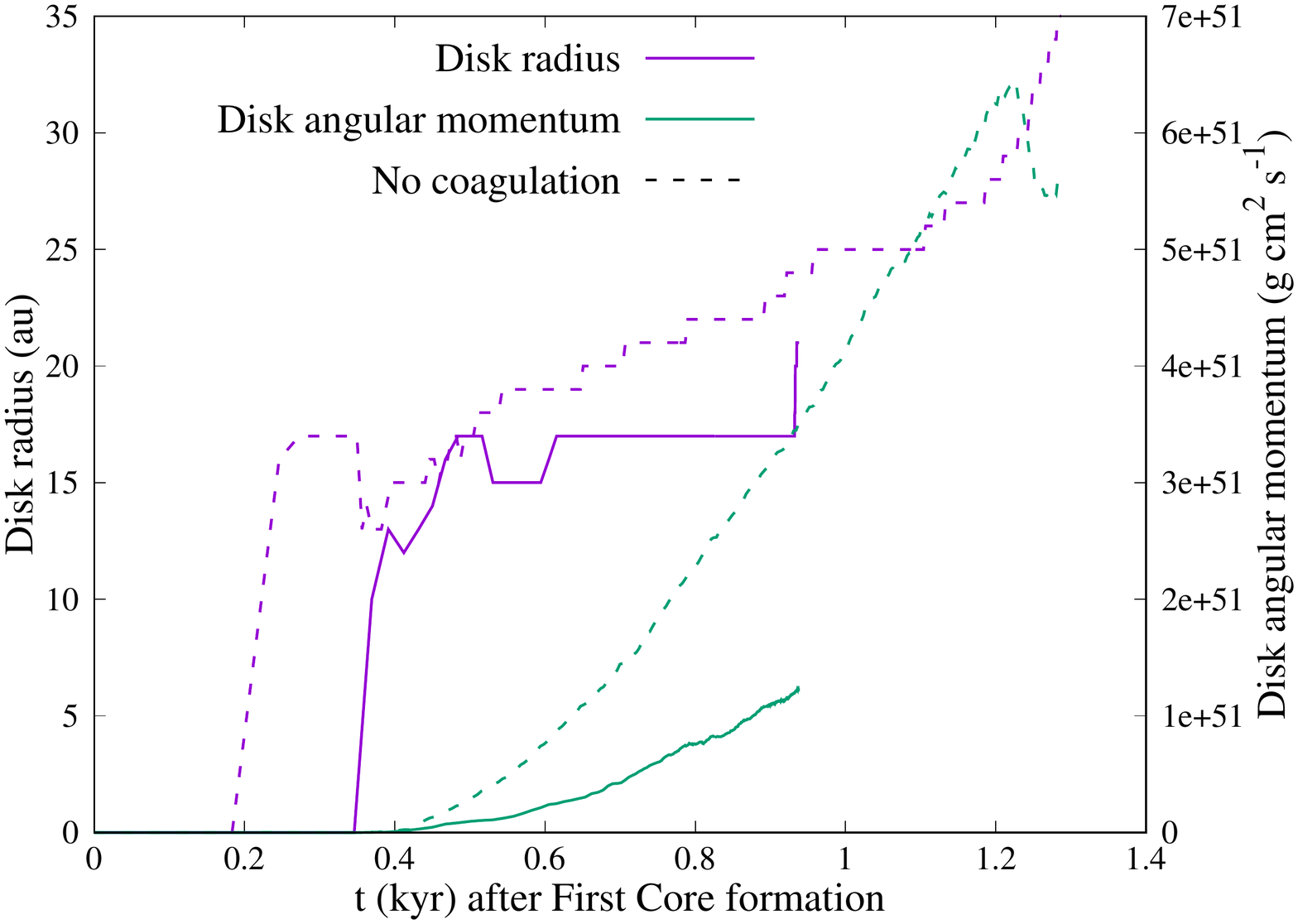}
  \caption{Disk size (purple, left axis) and angular momentum (green, right axis) as a function of time after the formation of the first Larson core. Solid lines represent simulation C-3 with coagulation, and dashed lines are simulation NC-3 without coagulation.}
  \label{FigDisksize}
\end{center}
\end{figure}

This difference in regime can be quantified by the ambipolar Elsasser number $\mathrm{Am}= B^2/(\rho \eta_\mathrm{AD} \Omega)$. In regions where $\mathrm{Am}<1$, the ambipolar diffusion has a significant impact on the dynamics of the gas. The bottom panel of Figure \ref{FigAveresist} shows the radial profile of the ambipolar Elsasser number in runs C-3 and NC-3, azimuthally averaged in the mid-plane. The higher resistivity in run NC-3 results in  $\mathrm{Am}<1$ in the inner $\sim 12$ au, indicating active ambipolar diffusion and magnetic field dissipation, while $\mathrm{Am} \gtrsim 10^4$ for C-3 over the same radial range, indicating weak ambipolar diffusion. Figure \ref{FigDisksize} compares the disk size and angular momentum in both simulations, further confirming the lower magnetic braking in run NC-3. 
A second effect of the weaker magnetic forces from the stronger ambipolar diffusion in run NC-3 is the absence of outflow at this stage of evolution, as shown in the lower panel of Figure \ref{FigMapsnocoag}.
     
The discrepancy of resistivity values between actual coagulation and methods simply redistributing the mass to the large-mass-end of the distribution originates from the lack of very large grains ($>10$ $\mu$m) in the latter. In both cases, removing the small grains decreases the electron absorption by dust, and therefore decreases the Ohmic resistivity. However, the ambipolar resistivity is controlled by the relative abundance of ions and charged grains in the gas. Although the small grain removal method barely changes the abundance of ions, the dominant positive charge carriers, it reduces significantly the charged grain population, leading to an increase in resistivity at low density. At high density, both with a standard MRN and a truncated-MRN distribution, the grains become the dominant charge carriers and the abundance of ions decreases (see the dashed lines in the top panel of Figure \ref{FigAveresist}). That does not happen with coagulation because the number of grains decreases significantly, hence, leading to a higher number of ions and a lower resistivity than without proper coagulation. This kind of method without larger grain creation is therefore inappropriate for emulating coagulation alone at a low cost for non-ideal MHD calculations. At later times, at densities $\rho > 10^{-12}$ g cm$^{-3}$, \citet{2023MNRAS.518.3326L} showed that the replenishment of small grains by fragmentation would lead to an increase in both Ohmic and ambipolar resistivities.

\section{Discussion and caveats} \label{sec:discussion}

\subsection{Grain growth}

As explained in previous sections and displayed in Figures \ref{FigChivsnhSimu} and \ref{FigAmaxvsnh}, initial gas conditions have little to no influence on grain coagulation for later stages and coagulation is ineffective in the envelope for growing large grains. This reinforces the idea that the system forgets the initial conditions at the formation of the first hydrostatic core and the disk \citep{2017A&A...598A.116V}. Consequently, calculations such as the ones presented in this work may provide standard initial dust grain size distributions for studies of protoplanetary disk evolution. Although other coagulation kernels affect the size distribution in different ways (see Section \ref{sec:kernels}), it is certain that even young protoplanetary disks contain grains that are significantly larger than those in the classical MRN distribution.
This has important implications for the dynamics of grains in the disk. Larger grains couple differently to the gas and may trigger the streaming instability \citep{2005ApJ...620..459Y,2007ApJ...662..627J,2017A&A...606A..80Y}, which is an early step toward planet formation. Although this regime is not reached in our simulations, the fast growth of the grains in circumstellar disks could predict an early onset for this process.

\subsection{Large grains in envelopes and dust diffusion}\label{sec:largegrainsenvelope}

Although grain coagulation is negligible in the envelope of our simulations, large grain signatures in envelopes have been observed. \citet{2019A&A...632A...5G}, for example, report "low and varying dust emissivity indices" at the envelope scale for some Class 0 and Class I protostars. This could be due to the presence of mm-size grains in low numbers. The time-scale to form such large grains in the envelope and cold ISM cores is large ($> 100$ Myr), so we exclude the possibility of early coagulation. Similarly, \citet{2019MNRAS.488.4897V} reported that their synthetic polarization observations of young protostellar envelopes in RAMSES calculations require grains larger than $10$ $\mu$m in order to be consistent with observations, which is also inconsistent with our findings and those of recent studies \citep[][for the latest ones]{2022ApJ...940..188S,2023MNRAS.518.3326L}.

The aerodynamic properties of the grains may cause differential velocities between grain populations and the gas, leading to varying dust-to-gas ratio throughout the cloud \citep{2020A&A...641A.112L}. We note, however, that (as these authors have found) dust diffusion will start to play a significant role only for very large grains of a few hundred microns. Generally, large grains tend to accumulate in higher density regions. \citet{2021ApJ...920L..35T} found that this dust diffusion can lead to what they call an ash fall phenomenon, in which large coagulated grains (up to millimeter-size) from the disk are ejected by an outflow, then decouple from the gas and fall back in the envelope. This process may explain the observations of low spectral index in Class 0 envelopes, as the outflow fuels the envelope with large grains formed in the central region. 
Eventually, those ejected grains circle back to the outer edge of the disk, enriching the large-end of the size distribution. In our case, the disk wind is fueled by the upper layers of the pseudo-disk, in which dust grows only moderately. However, dust-to-gas ratio enhancement in this region due to the grain differential velocities may lead to an accelerated growth.
Further studies are needed to investigate this discrepancy about the size of grains in envelopes between observations and theory.

\subsection{Coagulation in the pseudo-disk}
Grains in our simulations seem to grow faster than in \citet{2022MNRAS.514.2145B} despite their use of the same turbulent kernel as in our work. The peaks of the distributions in that work reach a few microns at a maximum density of $\nh=10^{13}$ cm$^{-3}$ (see their Figure 7), while in our simulations, the peak exceeds $100$ $\mu$m at a lower maximum density of 
$\nh \approx 3 \times 10^{12}$ cm$^{-3}$ (or $\rho \approx 10^{-11}$ g cm$^{-3}$; see our Figure \ref{FigAmaxvsnh}). That discrepancy is mainly due to a different modeling of the Reynolds number to calculate the turbulent coagulation kernel. They assumed a constant value of $\mathrm{Re}=10^8$, while we used \citep{2009A&A...502..845O,2020A&A...643A..17G}
\begin{equation}
    \mathrm{Re} = 6.7\times 10^{7} \left( \frac{\nh}{10^5~\mathrm{cm}^{-3}}\right)^{\frac{1}{2}}.
\end{equation}
Hence, in the central regions, our Reynolds number can be larger by three orders of magnitude. Their grains are then stuck in the tightly-coupled regime where relative grain velocities are lower than in the intermediate coupling-regime, which is more adapted to this situation (as demonstrated in Section 4.1 of paper I). The lower relative velocities result in lower coagulation rates and, therefore, a slower growth rate. The poor constraints on the value of the Reynolds number in protostellar environments therefore represents a source of uncertainty for grain growth by coagulation.

An additional explanation may involve an excess of growth in the pseudo-disk that forms in our simulations. The pseudo-disk is an over-density, typically denser than $\rho \approx 10^{-15}$ g cm$^{-3}$, created by the convergence of gas flowing along magnetic field lines \citep{GalliShu1993} that takes the shape of a disk perpendicular to the magnetic field, but is not supported against gravity. It is shown in the bottom panels of Figures \ref{FigMapscoag} and \ref{FigMapsnocoag}. 

Gas in the rotationally supported disk generally comes directly from the pseudo-disk and the overall efficiency of coagulation is mainly affected by the time spent in high-density regions such as the pseudo-disk. This is what appears in the bottom panel of Figure \ref{FigMapamax}, where there is a $\sim 30$ au-thick layer of large grains in the mid-plane. A passage of grains through the pseudo-disk would therefore provide an acceleration of coagulation in the early stage of star formation, even before arrival in the disk. That does not happen in \citet{2022MNRAS.514.2145B} since they do not consider magnetic fields, resulting in less-coagulated size distributions as grains enter the disk.

\subsection{Coagulation kernel}\label{sec:kernels}

Our coagulation methods works for every coagulation kernel where the environment variables and grain variables can be separated. One example of this is  the well-known turbulent kernel derived by \citet{2007A&A...466..413O} that we use in this work. 
This kernel is appropriate to calculate the growth of large grains, but it has several limitations. We assume a steady-state Kolmogorov turbulence spectrum and that the injection-scale of the turbulence is equal to the Jeans length corresponding to the local density (see Section 2.2 of paper I), which may lead to an overestimation of the grain collision rate.
Other kernels may have different effects on the size distribution of grains. \citet{2020A&A...643A..17G} showed that this turbulent kernel would increase the maximum size of the distribution, while a kernel derived from ambipolar diffusion, which creates a drift between charged and neutral grains, is efficient at removing small grains. This also happens in \citet{2022MNRAS.514.2145B} and \citet{2023MNRAS.518.3326L}, where combine several processes are combined, including Brownian motion and pressure gradients that generate drift velocities between grains and rapidly remove the smaller grains. These processes are not, however, efficient at producing larger grains without the help of the turbulent kernel. Contrary to fragmentation, the addition of those kernels would steepen the distribution at its lower end.

\section{Conclusions} \label{sec:conclusions}

We present here the results of numerical simulations of protostellar collapse with dust coagulation and self-consistent calculation of non-ideal MHD resistivities, using the methods detailed in \citet{2021A&A...649A..50M}. We performed four simulations, including three with coagulation and different collapse times and one without coagulation for reference. Here, are our main results.

\begin{itemize}
    \item[-] Coagulated size distributions retain some characteristics of the initial MRN distribution, in particular, the dominance of small grains in number and large grains in mass.

    \item[-] Dust coagulation is inefficient at growing larger grains in the envelope, even with long free-fall times. What matters is the time spent in high-density regions ($\rho > 10^{-15}$ g cm$^{-3}$) such as the pseudo-disk. Fragmentation can also be ignored in the cloud-collapse phase.
    
    \item[-] Grain growth is extremely rapid in the disk. The peak of the size distribution in mass, which is also the largest relevant grain size of the distribution, reaches 1~$\mu$m in the pseudo-disk and more than 100~$\mu$m in the inner disk only $10^3$~yr after its formation.
    
    \item[-] Grain sizes have a significant impact on non-ideal MHD resistivities. Coagulated grains result in resistivities up to four orders of magnitude lower than non-coagulated grains, with a significant impact on the dynamics of the disk. Simple redistribution approximations fail to capture this effect, as it occurs because of the growth of the largest grains.
\end{itemize}

Accounting for grain coagulation is therefore necessary in star formation and protoplanetary disk simulations, as grains grow rapidly to $\geq 100 \,\mu$m in radius in disks. The effects of grain growth on chemistry and radiative transfer will be explored in future papers.

\begin{acknowledgements}
P.M. acknowledges the financial support of the Kathryn W. Davis Postdoctoral Fellowship of the American Museum of Natural History and of the European Research Council (ERC) under the European Union’s Horizon 2020 research and innovation programme (ERC Starting Grant “Chemtrip”, grant agreement No 949278).  M.-M.M.L. was partly supported by US NSF grant AST18-15461.
U.L. acknowledges the financial support of the European Research Council (ERC) via the ERC Synergy Grant ECOGAL (grant 855130).
\end{acknowledgements}

\appendix

\section{Calculating the resistivities} \label{AppIonization}

\subsection{Solving for the ionization}

Let us assume a size distribution of grains divided into $N$ bins. We then have the following system of equations (see Paper I):

\begin{align}
Z_k =& \psi \tau_k + \frac{1-\epsilon^2 \Theta^2}{1+ \epsilon \Theta \alpha_k + \epsilon^2 \Theta^2},\label{EqsysZ}\\
\langle \tilde{J}(\tau_k) \rangle =& (1-\psi) + \frac{\frac{2}{\tau_k}\left[ \epsilon^2\Theta^2 + {\epsilon \Theta} \right]} {1 + \epsilon \Theta\alpha_k + \epsilon^2 \Theta^2},\label{EqsysJ}\\
\epsilon =& \frac{1-\psi}{\Theta\mathrm{e}^{\psi}}, \label{EqsysEps}\\
\ni=&  -\frac{1}{1-\epsilon} \sum_k n_k Z_k,\label{EqsysNi}\\
f(\psi)=& \frac{\svie \epsilon \ni^2}{\zeta \nh} + \frac{\ni v_\mathrm{i}}{\zeta \nh} \sum_k n_k \pi a_k^2\langle\tilde{J}(\tau_k)\rangle -1 = 0.\label{EqsysF}
\end{align}
Here, $\epsilon = \ne / \ni < 1$ is the ratio between the number of electrons and ions, $\psi$ is the ratio between the electric potential of the grains and the kinetic energy of electrons, $n_k$, $Z_k$, and $a_k$ represent the number density, the average charge and the radius of grains in bin $k$, $\zeta$ is the CR ionization rate, and $v_\mathrm{i}=[8 k_\mathrm{B}T/\pi\mu_\mathrm{i} m_\mathrm{H}]^{1/2}$ is the thermal velocity of ions, with $T$ the temperature, $\mu_\mathrm{i}$ the average atomic mass of ions, and $m_\mathrm{H}$ the mass of the proton. We also have $\Theta=s_\mathrm{e}[\mu_\mathrm{i}m_\mathrm{H}/m_\mathrm{e}]^{1/2}$, with $s_\mathrm{e}$ as the sticking probability of electrons on grains and $m_\mathrm{e}$ the electron mass. Then, $\langle\tilde{J}(\tau_k)\rangle$ represents the enhancement factor for ion recombination on grains, $\tau_k = a_k k_\mathrm{B}T/e^2$ is the reduced temperature of grains \citep{DraineSutin}, and $\alpha_k = [8/(\pi \tau_k)]^{1/2}$. Finally, $\svie = 2\times 10^{-7} [T/300]^{-1/2}$ is the collision rate between ions and electrons.

We solve the system of relevant equations (\ref{EqsysZ}-\ref{EqsysF}) for $\psi$ and find $\ni$, $\ne$, and $Z_k$ for all bins. A Newton-Raphson algorithm is described in details in the Appendix A of Paper I.

\subsection{The resistivities}

The collision time-scale of species $s$ with neutral H$_2$, the most abundant species in the gas, is given by:

\begin{equation}
    \tau_{s\mathrm{H}_2} = \frac{1}{\mathrm{a}_s\mathrm{He}} \frac{m_s+m_{\mathrm{H}_2}}{m_{\mathrm{H}_2}} \frac{1}{n_{\mathrm{H}_2}\langle \sigma \omega \rangle_s}.
\end{equation}
Here, $\mathrm{a}_s\mathrm{He}$ accounts for the collisions with He and is equal to 1.14 for ions, 1.16 for electrons, and 1.28 for grains \citep{DeschMouschovias}; $m_s$ and $m_{\mathrm{H}_2}$ are the mass of the species, $s,$ and the H$_2$ molecule; $n_{\mathrm{H}_2}$ represents the number density of H$_2$ (roughly equal to the density of the gas); $\langle \sigma \omega \rangle_s$ is the collision rate, taken from \citet{PintoGalli} for electrons and ions:
\begin{align}
    \langle \sigma \omega \rangle_\mathrm{e} &= 3.16 \times 10^{-11} v_\mathrm{rms,e}^{1.3},\\
    \langle \sigma \omega \rangle_\mathrm{i} &= 2.4 \times 10^{-9} v_\mathrm{rms,i}^{0.6},
\end{align}
where
\begin{equation}
    v_{\mathrm{rms,s}} = \left( \frac{8k_\mathrm{B} T}{\pi \mu_{s,\mathrm{H}_2}} \right)^{\frac{1}{2}},
\end{equation}
is in km s$^{-1}$, with $\mu_{s,\mathrm{H}_2}$ the reduced mass of species $s$ and H$_2$. These velocities have been calculated in a three-fluid formalism but we use them in our monofluid framework. See some related concerns in the appendix of \citet{2020ApJ...900..180M}.
For grains, the collision rate is given by \citep{DraineSutin,KunzMouschovias2009}:

\begin{equation}
    \langle \sigma \omega \rangle_k = \pi a_k^2 \left(\frac{8 k_\mathrm{B} T}{\pi m_{\mathrm{H}_2}}\right)^{\frac{1}{2}} \left(1+\left(\frac{\pi}{2\tau_k}\right)^{\frac{1}{2}}\right).
\end{equation}
We also define the conductivity, $\sigma_s$, and cyclotron frequency, $\omega_s$, of species $s$:
\begin{align}
    \sigma_s &= \frac{n_s q_s^2 \tau_{s\mathrm{H}_2}}{m_s}, \\
    \omega_s &= \frac{q_s B}{c m_s},
\end{align}
where $q_s$ is the electric charge of species $s$, $B$ the magnetic field strength and $c$ the speed of light.
The parallel, perpendicular, and Hall conductivities are expressed as:

\begin{align}
    \sigma_{||} & = \sum_s \sigma_s,\\
    \sigma_{\perp} & = \sum_s \frac{\sigma_s}{1+(\omega_s \tau_{s\mathrm{H}_2})^2},\\
    \sigma_\mathrm{H} & = \sum_s \frac{\sigma_s \omega_s \tau_{s\mathrm{H}_2}}{1+(\omega_s \tau_{s\mathrm{H}_2})^2}.\\
\end{align}
Finally, the Ohmic, Hall, and ambipolar resisitivities are defined as:

\begin{align}
    \eta_\mathrm{O} &= \frac{1}{\sigma_{||}},\\
    \eta_\mathrm{H} &= \frac{\sigma_\mathrm{H}}{\sigma_{\perp}^2 + \sigma_\mathrm{H}^2},\\
    \eta_\mathrm{A} &= \frac{\sigma_\mathrm{\perp}}{\sigma_{\perp}^2 + \sigma_\mathrm{H}^2} - \frac{1}{\sigma_{||}}.\\
\end{align}

\section{Coagulation and dust-to-gas ratio}\label{AppDusttogas}

Our coagulation model assumes that grains are perfectly coupled with the gas and well-mixed, so that the dust-to-gas mass ratio is constant. However, grains of different sizes have different aerodynamic properties and may experience a significant drift through the gas. This characteristic is usually determined by their Stokes number, that is, the ratio between the stopping time of the grain \citep{1924PhRv...23..710E} and the dynamical timescale of the system. In particular, \citet{2020A&A...641A.112L} showed strong variations in the dust-to-gas mass ratio during the protostellar collapse. The disk and first core tend to be dust-enriched while the envelope becomes dust-depleted. This has strong implication for the coagulation, as a higher grain density promotes collisions and speeds up their growth (conversely for a lower density).
In this section, we briefly explore a refinement to the coagulation model presented in Paper I. 

The general expression of the \citet{1916ZPhy...17..557S} equation, from which we derive the expression of $\chi$, is (Paper I, Equation 5):
\begin{equation}\label{EqDxdchi}
  \frac{\mathrm{d}X(a,t)}{\mathrm{d}t} = C g_\mathrm{local} n_\mathrm{H} I(a,X,t),
\end{equation}
where $C$ is a constant, $g_\mathrm{local}$ is a function of the local properties of the gas (density, temperature...), and
\begin{equation}
\begin{aligned}
  I(a,X,t)=& - \int_0^\infty h(m,m') X(m,t)X(m',t)\mathrm{d}m'\\
           & + \frac{1}{2} \int_0^m h(m-m',m') X(m-m',t)X(m',t)\mathrm{d}m'.
\end{aligned} \label{eq:IaXt}
\end{equation}
The relative number of grains of size, $a,$ at a given time, $t,$ is $X(a,t)=n_\mathrm{grain}/\nh$ where $n_\mathrm{grain}$ is the number density of grains. We can then write $X=d_\mathrm{g} x$, where $d_\mathrm{g}$ is the dust-to-gas ratio, and $x$ is the normalized relative number of grains. We can then rewrite Equation \ref{EqDxdchi}:
\begin{equation}\label{EqDxdchi_rewrite}
  \frac{\mathrm{d}x(a,t)}{\mathrm{d}t} = C g_\mathrm{local} n_\mathrm{H} d_\mathrm{g} I'(a,x,t),
\end{equation}
and include $d_\mathrm{g}$ in the definition of our coagulation variable, giving it the following form:
\begin{equation}
    \mathrm{d}\chi' = g_\mathrm{local} n_\mathrm{H} d_\mathrm{g} \mathrm{d}t.
\end{equation}
In our case, for the \citet{2007A&A...466..413O} kernel,
\begin{equation}
    \mathrm{d}\chi' = n_\mathrm{H}^{\frac{3}{4}} T^{-\frac{1}{4}} d_\mathrm{g} \mathrm{d}t.
\end{equation}
This expression means that the dust-to-gas ratio does not change the evolution path of the size distribution, only the coagulation speed. A dust-to-gas ratio, $N,$ times higher requires a $\chi$ value that is $N$ times lower to reach the same coagulated state. This new definition of $\chi'$ can be used in an environment with varying dust-to-gas ratio. In hydrodynamical simulations, it is possible to couple it with the drift of one grain size close to the peak of the mass density distribution, as allowed by the method proposed by \citet{2019A&A...626A..96L}.  

Although it is not yet possible to associate this coagulation method with the differential drift of grains of different sizes, this is a first step towards a self-consistent treatment of dust grains in hydrodynamics simulations. This method will lead to an overestimate of the small-to-larger grain ratio, but this is probably a decent approximation to get the total dust-to-gas ratio. As shown in \cite{2020A&A...641A.112L}, as long as the large grains dominate the mass in the distribution, they also dominate the differential gas and dust dynamics. In other words, most of the dust enrichment comes from the dynamics of large grains.

\section{The fragmentation barrier}\label{AppFragmentation}

\cite{2009A&A...502..845O} derived criteria for the fragmentation of dust aggregates. They define the rolling energy, $E_\mathrm{roll}$, that determines the energy needed to restructure the grain. In Section 4.2 of paper I, we misinterpreted their results by assuming the fragmentation would occur for a kinetic energy $E_{\rm kin} = 5 E_\mathrm{roll}$. The actual criterion is:
\begin{equation} \label{eq:fragcriterion}
    E_\mathrm{kin} > 5 N_\mathrm{tot} E_\mathrm{br},
\end{equation}
where $E_\mathrm{kin}$ is the kinetic energy, $E_\mathrm{br}$ is the breaking energy and $N_\mathrm{tot}$ is the total numbers of monomers composing the two colliding grains; $E_\mathrm{br}$ is defined by \citep{1997ApJ...480..647D}:
\begin{equation}
    E_\mathrm{br} = A_\mathrm{br} \gamma_\mathrm{grain}^{5/3} \frac{\left(a_0/2\right)^{4/3}}{\varepsilon^{2/3}},
\end{equation}
with $A_\mathrm{br}=2.8\times 10^{3}$, $\gamma_\mathrm{grain}$ as the surface energy density of the material, $a_0$ as the size of the monomers composing the grains, and $\varepsilon$ as the reduced elastic modulus. As in \citet{2009A&A...502..845O}, we adopt $a_0=0.1$ $\mu$m. Ice mantles on grains make them more resistant to fragmentation. Therefore, we assume bare silicates to obtain a lower limit for a fragmentation criterion. In this case, $\gamma_\mathrm{grain} = 25$ erg cm$^{-2}$ and $\varepsilon=2.8\times 10^{11}$ dyn cm$^{-2}$. 
The kinetic energy of two grains of mass $m$ and $m'$ is expressed as:
\begin{equation}
    E_\mathrm{kin} = \frac{1}{2} \frac{m m'}{m+m'} \Delta v^2,
\end{equation}
where $\Delta v$ is the relative velocity between the two grains.
This velocity is given by the kernel of \citet{2007A&A...466..413O} that we use in Paper I:
\begin{equation}
    \Delta v = \left( \frac{3}{\sqrt{8}} z_0 [k_\mathrm{B}G]^{\frac{1}{2}} \frac{\gamma \rho_\mathrm{s}}{\mu m_\mathrm{H}}\right)^{\frac{1}{2}} \nh^{-\frac{1}{4}} T^{-\frac{1}{4}} a^{\frac{1}{2}},
\end{equation}
where $z_0=2.97$, $k_\mathrm{B}$ and $G$ are the Boltzmann and gravitational constants, $\gamma=5/3$ the adiabatic index of the gas, $\rho_\mathrm{s}=2.3$ g cm$^{-3}$ the bulk density of the grains, $\mu=2.3$ the average atomic mass of the gas, $m_\mathrm{H}$ the proton mass, and $a$ the radius of the larger of the two grains. 
Assuming two identical grains and 
\begin{equation}
    m=\frac{4}{3}\pi \rho_\mathrm{s} a_0^3 N_\mathrm{tot},
\end{equation}
Equation \ref{eq:fragcriterion} becomes
\begin{equation}
    a > \frac{15 E_\mathrm{br}}{\pi \rho_\mathrm{s} a_0^3} \left( \frac{3}{\sqrt{8}} z_0 [k_\mathrm{B}G]^{\frac{1}{2}} \frac{\gamma \rho_\mathrm{s}}{\mu m_\mathrm{H}}\right)^{-1} \nh^{\frac{1}{2}} T^{\frac{1}{2}}.
\end{equation}
Replacing the variables with numeric values, we find
\begin{equation} \label{eq:frag}
    a > 826~\mu\mathrm{m} \left(\frac{\nh}{10^{10}~\mathrm{cm}^{-3}}\right)^{\frac{1}{2}}\left(\frac{T}{10~\mathrm{K}}\right)^{\frac{1}{2}}.
\end{equation}
This limit is much higher than the one derived in Paper I and we do not reach it in our simulations. 
The value of $\gamma_\mathrm{grain}$ we use is taken from \citet{2009A&A...502..845O}, but the surface energies of different materials can reach 10 to 100 times this value, resulting in a much higher estimate of the fragmentation threshold (as $E_\mathrm{br} \sim \gamma_\mathrm{grain}^{5/3}$). The value derived in Equation \ref{eq:frag} should then be considered very conservative.

\bibliographystyle{aa}
\bibliography{MaBiblio}

\end{document}